\documentclass[twocolumn]{aastex631}
\usepackage{amsmath}
\usepackage{longtable}
\begin{document}

\title{Revisiting the Perseus Cluster I: Resolving the Si/S/Ar/Ca ratios by Stellar Convection }

\shortauthors{Leung, Nomoto and Simionescu}
\shorttitle{Effects of Convective Processes on Si-group Production}

\author[0000-0002-4972-3803]{Shing-Chi Leung}

\affiliation{Department of Mathematics and Physics, SUNY Polytechnic Institute, 100 Seymour Road, Utica, NY 13502, USA}

\author[0000-0001-9553-0685]{Ken'ichi Nomoto}

\affiliation{Kavli Institute for the Physics and 
Mathematics of the Universe (WPI), The University 
of Tokyo Institutes for Advanced Study, The 
University of Tokyo, Kashiwa, Chiba 277-8583, Japan}

\author[0000-0002-9714-3862]{Aurora Simionescu}

\affiliation{SRON Netherlands Institute for Space Research, Niels Bohrweg 4, 2333 CA Leiden, The Netherlands}

\affiliation{Kavli Institute for the Physics and 
Mathematics of the Universe (WPI), The University 
of Tokyo Institutes for Advanced Study, The 
University of Tokyo, Kashiwa, Chiba 277-8583, Japan}

\affiliation{Leiden Observatory, Leiden University, PO Box 9513, 2300 RA Leiden, The Netherlands}

\correspondingauthor{Shing-Chi Leung}
\email{leungs@sunypoly.edu}

\date{\today}

\newcommand{\red}[1]{\textcolor{red}{#1}}

\submitjournal{ApJ}
\received{Jul 7 2024}
\revised{Jul 17 2025}
\accepted{Jul 24 2025}
\published{Sep 9 2025}

\begin{abstract}

Chemical abundance measurements from stars in the Milky Way to the intragalactic medium in the Perseus Cluster have challenged the spherical explosion models. Models in the literature cannot closely match the observed element ratios, where Si, S are overproduced and Ar, Ca are underproduced. In this article, we explore the impact of the model parameters during the evolution of massive stars on the final explosive nucleosynthesis. We investigate the effects of a parametrized model of the convective process, including the mixing length parameter and the semi-convection parameter, on the production of Si-group elements. 
We search for the value pair that can reduce the discrepancy in the models. 
We conclude that a mixing length parameter of 2.2 and semi-convection parameter of 0.03 are required to fit these criteria. Using this updated value pair, we compute a sequence of massive star models from $M_{\rm ZAMS} = $ 15 -- 40 $M_{\odot}$. 
The high resolution data from future observations such as XRISM will provide further details on less constrained processes in stellar evolution and supernova explosion. Future comparison with supernova models of various progenitor metallicity will further shed light on the supernova population and their relative rates on cosmological scales.

\end{abstract}





\section{Introduction}

\subsection{Massive Star Explosions}

Massive star explosion models are connected to various important fields in astrophysics: (1) their explosions correspond to the two large subclasses of Type Ib/c and Type II supernovae \citep[e.g., ][]{Wheeler1990TypeIReview, Nomoto1993SN1993J, Nomoto1994SN1994I, Hillier2019SNII}, (2) they are the main contributors of the alpha-elements (C, O, Ne, Mg) and Si-group elements (from Mg to Ca) \citep{Arnett1996}, (3) their explosion picture reveals the interaction between neutrinos and nuclear matter \citep{Janka2012CCSNReview} and (4) they play roles in the galactic chemical evolution, especially in early phase \citep[see recent works, e.g., ][]{Kobayashi2020GCEFull}. Nucleosynthesis of stars allows us to trace their evolution through the perspective of chemical elements \citep[see reviews, e.g., ][]{Hoyle1956, Burbidge1957ElementsStars, Woosley2002MassiveStarReview, Nomoto2013NucleoReview, Thielemann2019NucleoFingerprint}. In particular, the chemical composition of very or extremely metal-poor stars are the ideal probe for explosions of a single or a few massive stars \citep{Hartwig2023MLMPS}. 

Neutrino heating is the main driver for massive star explosions. The Standing Accretion Shock Instability \citep{Foglizzo2002SASI, Fernandez2010SASI, Fernandez2014SASI} allows the stalled shock to continue to grow and trigger the explosion. However, current models show that the explosion robustness depends on the dimensionality, initial profiles and microphysics \citep{Burrows2013CCSNReview, Janka2016Review, Burrows2018CCSNInputPhys}. Spherical core-collapse explosions with neutrino heating fail to explode massive stars robustly \citep[e.g., ][]{Mezzacappa2001SphereicalFail, Liebendoerfer2001GRCCSNFail}.

Under such difficulties of modeling the core-collapse supernovae (CCSNe), approximate simulations have been tried to obtain explosion models for comparison with observational data. Here an empirically parameterized energy is given at inner mass cut. 
Works focusing on the ejecta dynamics and nucleosynthesis typically treat the energy source as a thermal bomb \citep[e.g., ][]{Umeda2002PopIIICCSN} or a piston \citep[e.g., ][]{Woosley1995CCSN}. These physical pictures aim at providing the stellar envelope the necessary energy and momentum to expand and to be ejected. More recent treatments include parametrized models of neutrino heating \citep{Janka1996PTOBCode, Sukhbold2016CCSN}. These different prescriptions provide a diversity of individual massive star yields \citep{Woosley1995CCSN, Limongi2003CCSN, Tominaga2007, Sukhbold2016CCSN}.

Massive star models have attracted interest for comparison with nearby CCSN explosions. Their spectra provide first-hand constraints on the explosion process, and hence the thermodynamics history in these supernovae, for instance SN 1987A \citep{Hashimoto1989SN1987A} and Cassiopeia A \citep{Siegert2015CasATi44}. In recent years, the rapid discoveries of extremely metal-poor stars (EMPS) \citep[such as the catalogue from LAMOST, ][]{Li2018LAMOST10000} and the centralized storage database \citep[such as SAGA, ][]{Suda2008SAGA} provide an alternative approach to constrain supernova explosions. Some unusual ones inspire the development of the theoretical models. For example, the zinc-enriched EMPS HE 1424-0241 indicates the possibility of aspherical explosion \citep{Tominaga2009Jet, Leung2023Jet1}. The strong contrast in the odd-number charged elements over even-number charges elements in some very metal-poor stars (VMPS) such as LAMOST J1010+2358 could indicate its very massive nature, suggesting the possibility of pair-instability supernova \citep{Xing2023LAMOST}. 

Besides individual stars, systems such as galaxies and galactic clusters provide a global view to the supernova explosion as an entity. The galaxies as the birth place of stars and the repository of the stellar and supernova ejection encode how generations of stars evolve and how metals are accumulated and recycled through the cosmic history. Through the trends of individual elements over metallicity [Fe/H], we may derive constraints on supernova yields and their individual rates \citep{Timmes1995GCE, Kobayashi2020GCE, Nomoto2013NucleoReview, Mateteucci2001GCE}. The hot gases in the galactic clusters, ejected from billions of supernovae from hundred to thousands of galaxies, may serve a similar role to test supernova models \citep{Mernier2017ICM, Mernier2022GalClusterBook}. It is surprising to note that the chemical composition of the hot gas is very close to solar composition in various galactic clusters (e.g., Centarius, M87, Perseus Cluster) \citep{Mernier2018Solar}, potentially suggesting a universality of the enrichment process. 

\subsection{Motivation}

Detailed chemical abundance measurements are made available across the cosmos and across multiple wavelengths – for Milky Way stars from optical spectroscopic surveys like APOGEE \citep{APOGEE2018DR14} and GALAH2 \citep{Buder2018GALAH2}, for distant galaxies (from JWST), for the intracluster medium (ICM) and supernova remnant (SNR; from Hitomi and now XRISM). These missions brought us precise chemical abundance patterns which allow us to quantitatively distinguish physics in stellar evolution and supernova explosion. While the models in literature generally follow the observed patterns reasonably well, they cannot reproduce many of the details now being constrained from the new data, especially $\alpha$-chain elements. 

A representative example is the galactic chemical evolution. Through mapping the element ratios as functions of metallicity [Fe/H] from stellar surveys in the Milky Way, \cite{Griffith2021Bulge} reported challenges in reproducing the O/Mg ratios in Milky Way stars (from GALAH2 and APOGEE); since these elements are almost exclusively contributed by massive star explosions, the observed discrepancies can be viewed as empirical clues that aspects of massive star evolution or explosion physics are still missing from the nucleosynthesis models. 

Another prominent example is the ICM of the Perseus Cluster. The mismatch was hinted in early measurement by XMM-Newton \citep{dePlaa2007Cluster,Mernier2016ClusterYield} and confirmed by the detailed Hitomi observations of Perseus. Interestingly, similar residuals are in fact seen in some Galactic Chemical Evolution models for instance, it appears that less Si and S and more Ar and Ca are detected compared to the predictions of publicly available supernova yields. 
With more than 400 galaxies involved, the Intracluster Medium (ICM) in the Perseus Cluster is a relic of over billions of supernova ejecta across the cosmic history. The substantial scale could average out peculiarity in individual stars or even individual galaxies, making its chemical composition an ideal laboratory for studying how stars and supernovae synthesize chemical elements from a very general approach. This approach has been used for constraining relative rates of CCSN and Type Ia Supernovae (SNe Ia) \citep{Tsujimoto1995SNRate}. 


As the brightest galaxy cluster in the sky, Perseus has long served as a quintessential target for studying many processes that shape the Intracluster Medium (ICM), including its chemical enrichment pattern. Using early observations with moderate spectral resolution CCD detectors, it was already reported that the composition of the ICM (looking at various elements relative to Fe) was roughly consistent with Solar \citep[e.g., ][]{Tamura2009Perseus, Matsushita2013Perseus}, and spatially uniform \citep{Mernier2017ICM}. The Hitomi X-ray telescope \citep{Takahashi2016Hitomi} offered a much higher spectral resolution view of the Perseus Cluster core, enabling remarkably precise abundance measurements to be derived for Si-group elements (Si, S, Ar, Ca) as well as Fe-group elements (Cr, Mn, Ni) \citep{Hitomi2017Nat}. This revealed that the composition of the ICM is indeed in strikingly good agreement with the Solar ratios -- yet, difficult to reconcile with existing supernova yield models.

Some recent works performed extensive comparison between models in the literature and the new data. \cite{Simionescu2019Perseus} performed a detailed comparison of the abundances measured in the Perseus Cluster by Hitomi with numerical models for CCSNe reported in \cite{Nomoto2006Yields, Sukhbold2016CCSN} and SNe Ia reported in \cite{Iwamoto1999SNIa, Pakmor2012Merger, Nomoto2018Review, Leung2018Chand, Shen2018SubChand}. However, no combination of CCSN + SN Ia models can fully match the observed enrichment pattern. Specifically, models tend to overproduce Si and S, and underproduce Ar and Ca (See their Figure 7). These 4 elements are approximately equally contributed by the two classes of supernovae. Meanwhile, the combination of supernova models also overproduce Cr and underproduce Mn, which are mostly contributed by SNe Ia.  

In view of the mismatch, it becomes interesting to revisit the Perseus Cluster and examine how we can use the abundances of the Si-group elements as constraints to make our massive star models more realistic. The production of Si-group elements are sensitive to a number of factors, including nuclear physics (e.g., reaction rate of $^{12}$C$(\alpha,\gamma)^{16}$O), and semi-convection, treatment of convective and semi-convection overshooting mixing during the final O-burning phase \citep[see e.g.,][]{Thielemann1985MasStarSi, Weaver1993Rate, Woosley1995CCSN}. 
The Perseus Cluster is known to be very close to the Solar composition \citep{Mernier2018Solar}. Type Ia supernovae yields are also close to the Solar composition \citep[e.g.,][]{Nomoto2017SNIa, Nomoto2018Review}. Meanwhile, noted in \cite{Simionescu2019Perseus}, the massive star explosion models do not closely match the Solar composition with a remarkable underproduction in Ar and Ca. Therefore, in this project, we seek for models with enhanced Ar and Ca production such that Si-group elements agree with the pattern in Solar composition, while maintaining other elemental production unchanged, including Cr, Mn and Ni. 

We focus on the role of convective mixing parameters as these parameters are subject to multi-dimensional processes which are not completely resolved. This includes the mixing length parameter $\alpha$ and semi-convection parameter $\alpha_{\rm SC}$. These two parameters are frequently used for matching observed stellar properties. The parameter $\alpha$ is used to describe the efficiency of convective mixing \citep{BohmVitense1958MLT}. The calibration with the Sun gives a value $\alpha = 1.786$ \citep{Sonoi2019MLTalpha}. However, extension to other stars suggest that the parameter could be a function of metallicity, surface gravity and the effective temperature \citep{Viani2018alphafit}. For example, in \cite{Joyce2020Betelgeuse} the fitting of Betelgeuse luminosity, effective temperature and oscillation period suggests a higher value of $\alpha = 2.1$. This implies that the actual value could be system dependent. We therefore choose this as one potential parameter for affect the Si-group element production. Similarly, the parameter $\alpha_{\rm SC}$ determines the diffusion coefficient in the semi-convection regime \citep{Langer1983SC}. This parameter controls the extent of the semi-convection zone in H- and He-burning. It has a high uncertainty in its value, where a wide range of $\alpha_{\rm SC} \in (0.01,300)$ is studied. A value of $\sim 1$ is found for reproducing the demography of stars in the Small Magellanic Cloud \citep{Schootemeijer2019alphaSC}. This parameter is also system dependent where the required value decreased with mass \citep{Silva2012alphaSC}. In this work, we assume both parameters are free parameter. We remark that these parameters, unlike changing directly the nuclear reaction rates, do not directly change the nucleosynthetic yield. Instead, it changes the Si-group element production by affect the initial composition due to convective mixing.

In this article, we first introduce in Section \ref{sec:method} the codes that we use for the stellar evolution, core-collapse simulations and explosive nucleosynthesis. In Section \ref{sec:mixing} we use the 20 $M_{\odot}$ model to demonstrate how the mixing parameters affect the stellar evolution and the distribution of Si-group elements in the pre-collapse massive star models. In Section \ref{sec:post-collapse} we present how these models undergo core-collapse explosion and how the final chemical abundance is affected by these stellar evolution processes. Then, we discuss the effects of the convective mixing parameters to the nucleosynthetic yield of the reference model and extend our computation to massive star models to 15 -- 40 $M_{\odot}$ stars in Section \ref{sec:catalogue}. 
Finally, in Section \ref{sec:discussion}, we discuss the caveats of our models and compare our massive star yields with some representative models in the literature. Finally we give our conclusion. 

\section{Method}
\label{sec:method}

For the stellar evolutionary models, we use the open-source code MESA \citep[Modules for the Experiments in Stellar Astrophysics,][version 8118]{Paxton2011MESA, Paxton2013MESA, Paxton2015MESA, Paxton2018MESA, Paxton2019MESA}. The code solves for the quasi-hydrostatic stellar structure assuming spherical symmetry. The models assume local thermodynamical equilibrium with nuclear reactions as the main energy source. The \texttt{Helmholtz} equation of state \citep{Timmes1999Helmholtz, Timmes2000Helmholtz} is responsible for the majority parts of the stars, which contains contributions from electron gas of arbitrary relativistic and degeneracy levels, nuclei in the form of an ideal gas, photon gas in blackbody distribution and $e^- - e^+$ pairs. In this work, we use the 21-isotope network which includes the hydrogen burning network, $\alpha$-chain network and Fe isotopes. Neutron-rich isotopes are represented by the isotope $^{56}$Cr. 

To explore the dependence of Si-group elements on the stellar process, we focus on two parameters: (1) the mixing length parameter $\alpha$ and the semi-convection parameter $\alpha_{\rm SC}$\footnote{The configuration files are available on Zenodo with the DOI:10.5281/zenodo.12667505. We also refer readers to \cite{Leung2021Wave, Leung2021SN2018gep} for similar construction of massive stars.} by the following configuration:
\begin{verbatim}
    mixing_length_alpha = 2.2
    alpha_semiconvection = 0.03
\end{verbatim}
After H-burning in the main-sequence, the H-envelope plays a minimum role to the later dynamics of other inner cores due to its extended structure away from the core. To facilitate the computation, once the star completes its H-burning in the main-sequence, the H-envelope is removed. Then we proceed to model the advanced evolution. The stellar evolution model stops when the Fe-core begins its gravitational collapse. 

To compute the post-collapse phase of the massive stars, we first transfer the numerical model to the one-dimensional explosion code \citep{Tominaga2007}. The code solves the reactive Lagrangian hydrodynamics in spherical symmetry with the $\alpha$-chain network. The collapsed Fe-core is represented by the inner mass cut $M_{\rm cut}$ and a thermal bomb $E_{\rm dep}$. The simulations are terminated when the ejecta density is low enough such that no substantial exothermic reactions take place. For later computation of explosive nucleosynthesis, the thermodynamics history of each fluid element is also recorded. 

After the explosive hydrodynamics, we pass the thermodynamics history to the nuclear reaction network \texttt{torch} \citep{Timmes1999Torch} for computing the detailed nucleosynthesis. We use the 495-isotope network which includes isotopes from $^{1}$H to $^{93}$Tc. To compute the stable isotopes, we continue to process the ejecta for further $10^6$ years so that all short-lived isotopes (e.g., $^{56, 57, 59}$Ni) are all decayed to their stable daughter nuclei. For comparison with the solar abundance we refer to the mass fraction from \cite{Asplund2009Solar}.

\section{Effect of Mixing Parameters}
\label{sec:mixing}

\begin{table*}[]
    \caption{The model parameters for the stellar evolution models studied in this work with the global parameters of the models. The ZAMS mass $M_{\rm ZAMS}$, He-core mass $M_{\rm He,C}$, C+O core mass $M_{\rm C+O,C}$, Si-core mass $M_{\rm Si,C}$ and Fe-core mass $M_{\rm Fe,C}$ are in units of $M_{\odot}$. The total masses of individual elements ($M$($^{16}$O), $M$($^{28}$Si), $M$($^{32}$S), $M$($^{36}$Ar) and $M$($^{40}$Ca)) are also in units of $M_{\odot}$.}
    \centering
    \begin{tabular}{c c c c c c c c c c c c c}
        \hline
         Model & $M_{\rm ZAMS}$ & $\alpha$ & $\alpha_{\rm SC}$ & $M_{\rm fin}$ & $M_{\rm C+O,C}$ & $M_{\rm Si,C}$ & $M_{\rm Fe,C}$ & $M$($^{16}$O) & $M$($^{28}$Si) & $M$($^{32}$S) & $M$($^{36}$Ar) & $M$($^{40}$Ca) \\ 
         Unit & $M_{\odot}$ & none & none & $M_{\odot}$ & $M_{\odot}$ & $M_{\odot}$ & $M_{\odot}$ & $M_{\odot}$ & $M_{\odot}$ & $M_{\odot}$ & $10^{-2}~M_{\odot}$ & $10^{-2}~M_{\odot}$\\ \hline
         M20A15S01 & 20 & 0.15 & 0.01 & 5.50 & 3.83 & 2.01 & 1.50 & 1.02 & 0.21 & 0.10 & 1.71 & 1.17 \\
         M20A18S01 & 20 & 0.18 & 0.01 & 5.47 & 3.81 & 2.54 & 1.44 & 0.98 & 0.29 & 0.16 & 3.06 & 2.00 \\ \hline
         M20A20S01 & 20 & 0.20 & 0.01 & 5.54 & 3.89 & 3.38 & 1.37 & 1.27 & 0.37 & 0.11 & 1.81 & 1.44 \\
         M20A20S03 & 20 & 0.20 & 0.02 & 5.51 & 3.85 & 2.12 & 1.44 & 0.96 & 0.26 & 0.17 & 3.07 & 2.77 \\
         M20A20S06 & 20 & 0.20 & 0.06 & 5.50 & 3.85 & 1.90 & 1.43 & 1.18 & 0.30 & 0.11 & 1.93 & 1.47 \\ 
         M20A20S10 & 20 & 0.20 & 0.10 & 5.50 & 3.86 & 2.05 & 1.46 & 1.14 & 0.12 & 0.04 & 1.17 & 1.34 \\ \hline
         M20A22S01 & 20 & 0.22 & 0.01 & 5.51 & 3.96 & 2.12 & 1.40 & 0.91 & 0.31 & 0.20 & 3.74 & 3.53 \\
         M20A22S03 & 20 & 0.22 & 0.03 & 5.51 & 3.86 & 2.74 & 1.41 & 0.99 & 0.31 & 0.19 & 3.73 & 2.77 \\
         \hline
    \end{tabular}
    
    \label{tab:models}   
\end{table*}

In Table \ref{tab:models} we list the massive star models studied in this work. The model name MXXAYYSZZ stands for the progenitor with a mass XX $M_{\odot}$, a mixing length parameter $\alpha = 0.1 \times YY$ and a semi-convection parameter $\alpha_{\rm SC} = 0.1 \times ZZ$. In that convention, M20A22S03 stands for a 20 $M_{\odot}$ model using $\alpha = 2.2$ and $\alpha_{\rm SC} = 0.03$. We vary $\alpha$ between 1.5 to 2.2 and $\alpha_{\rm SC} = 0.03$ between 0.01 - 0.30. In this section, we focus on the effect of these mixing parameters on the stellar evolutionary models. In the next section, we extend our calculation to various progenitor masses. 

From the table, by comparing models with varying $\alpha$ or $\alpha_{\rm SC}$. they do not show significant change in the nucleosynthetic pattern. An exception appears for M20A20S01 and M22A22S03. In these models, the Si-rich material has partially mixed with the convective layers in the O-layer. Therefore, Si becomes abundant at a much higher mass coordinate. Excluding the exceptional models, a higher $\alpha$ results in an observably higher production of Ar and Ca by about $\sim 50\%$. On the contrary, $\alpha_{\rm SC}$ has an opposite that when $\alpha_{\rm SC}$ is small, Ar and Ca production is high; the production decreases as $\alpha_{\rm SC}$ increases.

\begin{figure}
    \centering
    \includegraphics[width=8.5cm]{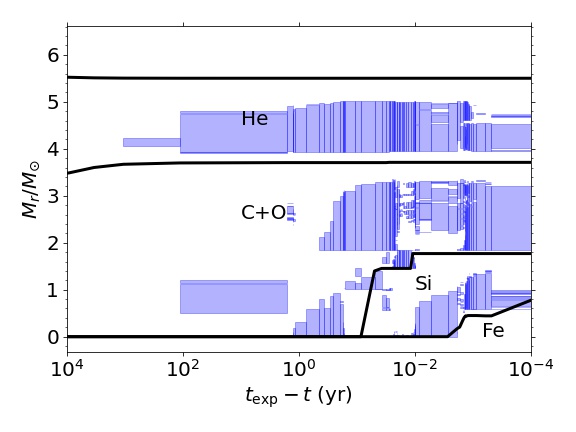}
    \includegraphics[width=8.5cm]{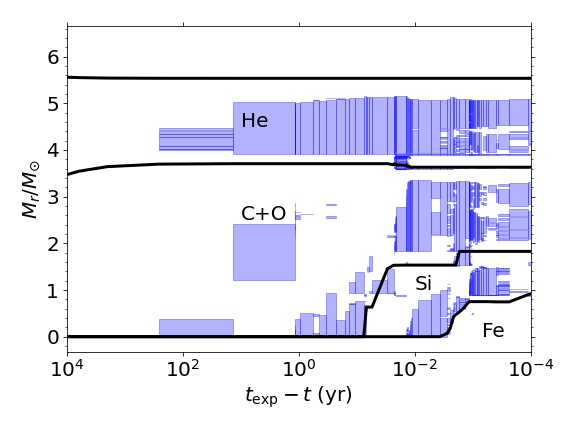}
    \caption{(top panel) The Kippenhahn diagram of M20A15S01. The lines from top to bottom correspond to the mass coordinates of the He, C+O, Si and Fe cores. (bottom panel) Same as the top panel but for the M20A20S01.}
    \label{fig:kipp}
\end{figure}

In Figure \ref{fig:kipp} we plot the Kippenhahn diagram\footnote{A Kippenhahn diagram plots the convection structure and various shell masses as a function of time $t^* = t_{\rm exp} - t$, measured from the onset of final collapse and explosion. It visualizes the mass loss, and the formation of various mass shells, and various nuclear burning phases.} of two contrasting models, M20A15S01 and M20A22S01, to distinguish the effect of $\alpha$ in the stellar evolution. A higher $\alpha$ leads to a more progressive Si-burning. In M20A15S01, the formation of Si core is prompt; while in M20A22S01, the Si layer mass grows continuously, resulting in a higher Si core mass in the end. Similar effects are observed for the Fe-core formation. Some of the convection history in the C+O layer is observed in M20A15S01 but not in M20A22S01. However, no significant changes are observed in the He layer. 

\begin{figure}
    \centering
    \includegraphics[width=8.5cm]{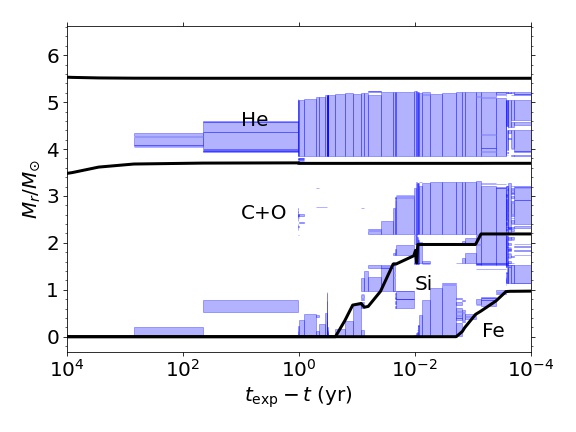}
    \includegraphics[width=8.5cm]{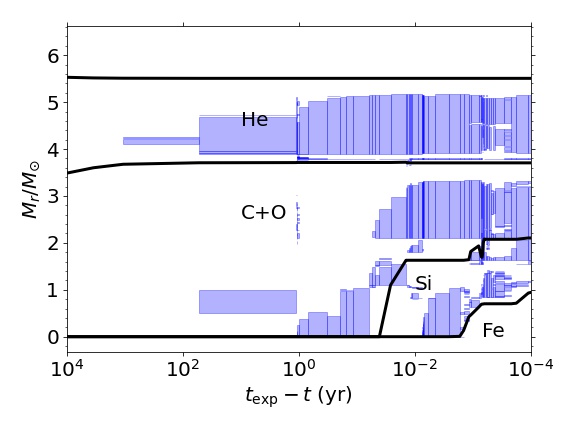}
    \caption{(top panel) The Kippenhahn diagram of M20A22S01. The lines from top to bottom correspond to the mass coordinates of the He, C+O, Si and Fe cores. (bottom panel) Same as the top panel but for the M20A22S03.}
    \label{fig:kipp2}
\end{figure}

In Figure \ref{fig:kipp2} we plot the Kippenhahn diagram of two other contrasting models, M20A20S01 and M20A22S03. The profiles show typical structure as in massive star evolution. In the last 100 years, He-shell burning creates persistent convection in the He-shell. C+O burning is triggered around 0.1 year before collapse and it builds up the Si core promptly. Meanwhile Si-burning forms the Fe core at $t^* \sim 10^{-3}$ yrs. 

M20A22S03 captures the typical feature when $\alpha_{\rm SC}$ is large. We observe that the strong convection (higher $\alpha$, $\alpha_{\rm SC}$) plays a role in creating a more extended Si-core during the final C-shell burning at $t^* \sim 10^{-3}$ yrs. In most other stellar evolution phases, no significant effects can be observed. 

M20A20S01 is the peculiar model. A major qualitative difference is the different Si core mass at $t^* \sim 10^{-3}$ yrs. The convection zone in the C+O layer penetrates into the Si layer. This suggests some additional mixing of the C+O-rich material that is later burnt to Si-group elements before its collapse. As a result, the diluted material leads to very contrasting fingerprint in the final chemical abundance pattern. In fact, we will show in later figure, that M20A20S01 is the peculiar model due to this convection zone interaction. 

\begin{figure}
    \centering
    \includegraphics[width=8.5cm]{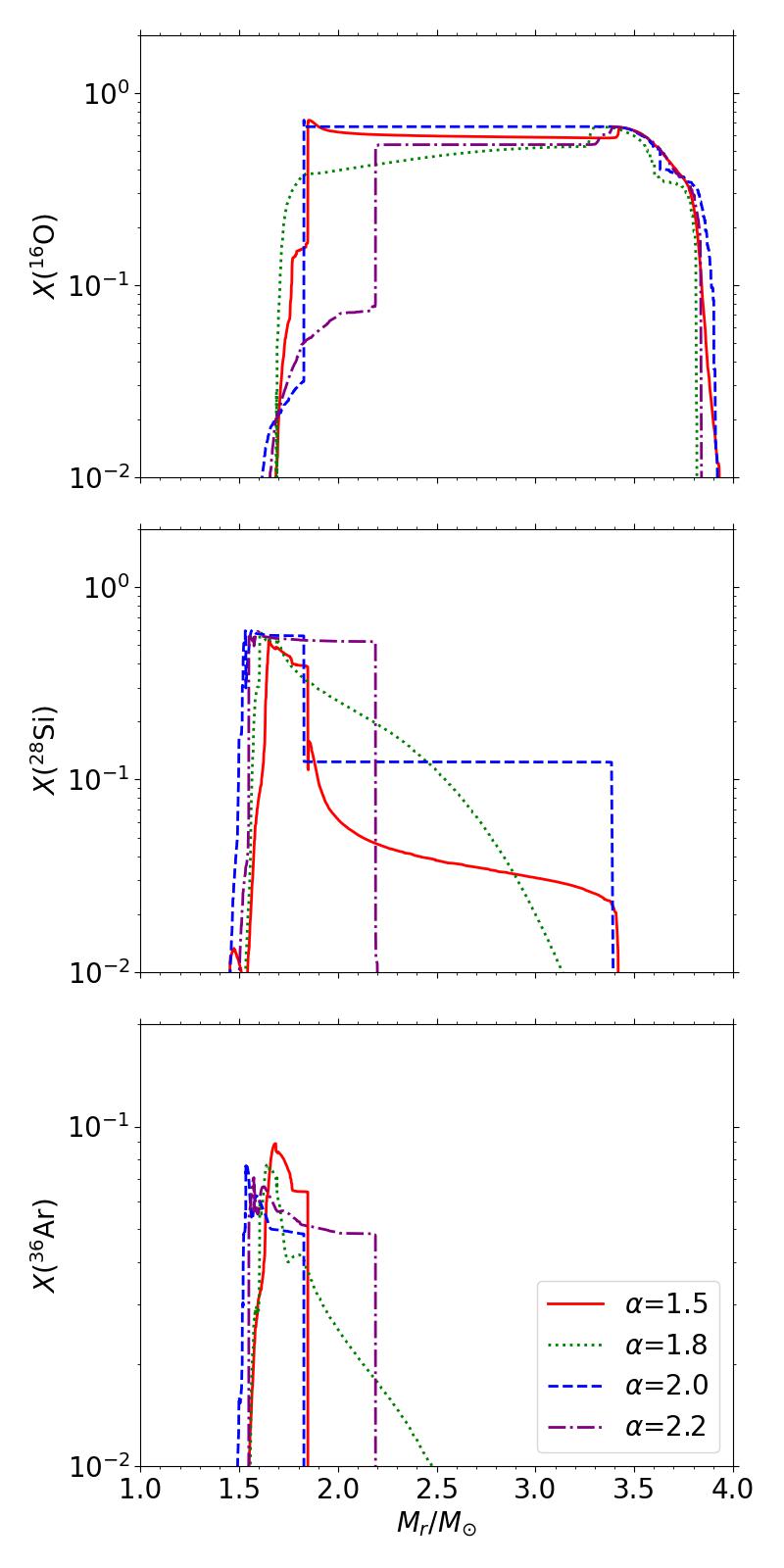}
    \caption{The pre-collapse isotopic abundance profiles for M20A15S01 (red solid line), M20A18S01 (green dotted line), M20A20S01 (blue dashed line), and M20A22S01 (purple dot-dashed line) for $^{16}$O, $^{28}$Si, and $^{36}$Ar, respectively.}
    \label{fig:precoll_profile_alpha}
\end{figure}

In Figure \ref{fig:precoll_profile_alpha} we plot the final profiles of the mass fractions $X$ of $^{16}$O, $^{28}$Si, and $^{36}$Ar for M20A15S01, M20A18S01, M20A20S01, and M20A22S01 to contrast the effects of $\alpha$ on the pre-collapse structure. The mixing length parameter has a highly non-linear role in the chemical structure of the ejecta. While there is no significant change in the zone in the plateau part, how the shell bridges to the next shell depends sensitively on the choice of $\alpha$. For $^{16}$O, for example, a high $\alpha$ can lead to the consumption of $^{16}$O in the lower boundary of the Si-core. We remark that M20A20S01 also shows an extended O-inner boundary, but the mixing leads to a higher $^{16}$O in the overall O-layer. For $^{28}$Si, the diffusion of the element appears only for $\alpha = $ 1.8 and 2.0, but not the other two cases. Finally, for $^{36}$Ar, a higher $\alpha$ leads to a higher mass fraction and more extended $^{36}$Ar in the Si-shell. The smaller size of Ar in M20A20S01 suggests that Ar formation is restricted after the convection zones merge. The shell merger can play an influencing role to the chemical structure. Demonstrated in \cite{Roberti2025ShellMerger}, the shell merger provides unique thermodynamics trajectory for fluid element in the Si-shell, which can facilitate the production of odd-number elements (e.g., K and Sc) compared to the model without any merger history. 

\begin{figure}
    \centering
    \includegraphics[width=8.5cm]{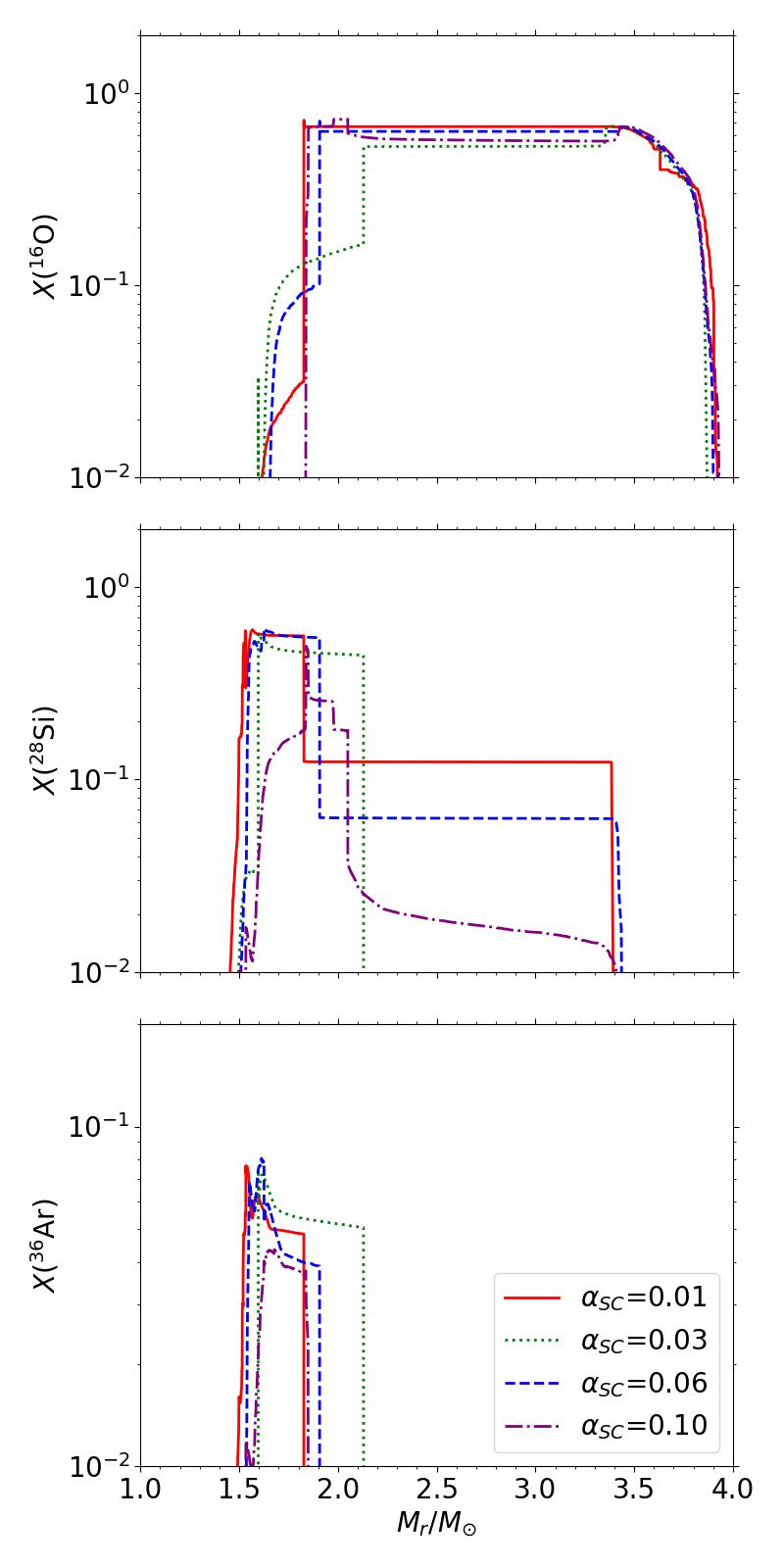}
    \caption{The pre-collapse abundance profile for M20A20S01 (red solid line), M20A20S03 (green dotted line), M20A20S06 (blue dashed line), and M20A20S10 (purple dot-dashed line) for $^{16}$O, $^{28}$Si and $^{36}$Ar respectively.}
    \label{fig:precoll_profile_fov}
\end{figure}

Similar to Figure \ref{fig:precoll_profile_alpha}, we plot in Figure \ref{fig:precoll_profile_fov} the pre-collapse profiles of the mass fractions of $^{16}$O, $^{28}$Si, and $^{36}$Ar for M20A20S01, M20A20S03, M20A20S06 and M20A20S10. These models examine the role of $\alpha_{\rm SC}$ on the final chemical production. By contrasting the latter three models, a high
$\alpha_{\rm SC}$ (S01-S10) leads to a more clear cut between the O and Si layer, resulting in more overall O in the star.  At the same time, a high $\alpha_{\rm SC}$ suppresses the formation of Si, leading to more step-like structure in the profile. The same parameter also reduces the production of Ar as shown by the reduced size of the Ar-rich zone and its overall abundance. 

\begin{figure}
    \centering
    \includegraphics[width=8.5cm]{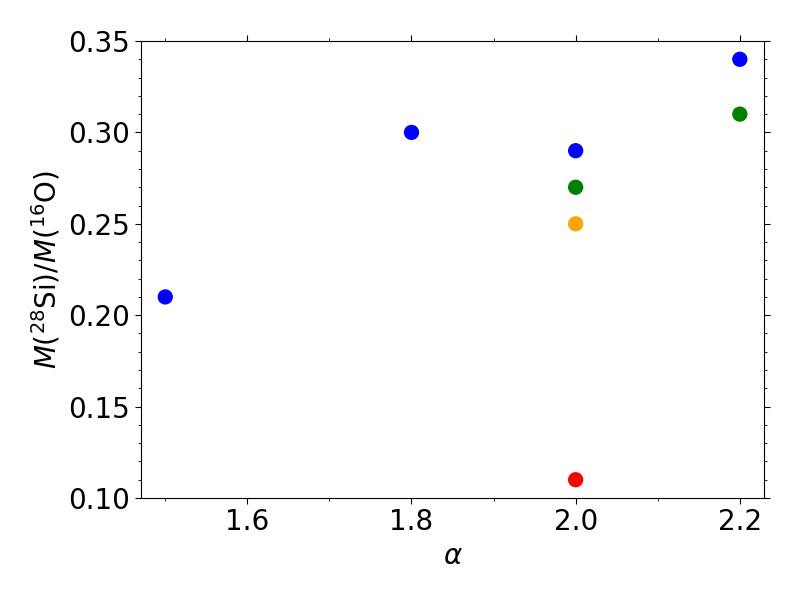}
    \includegraphics[width=8.5cm]{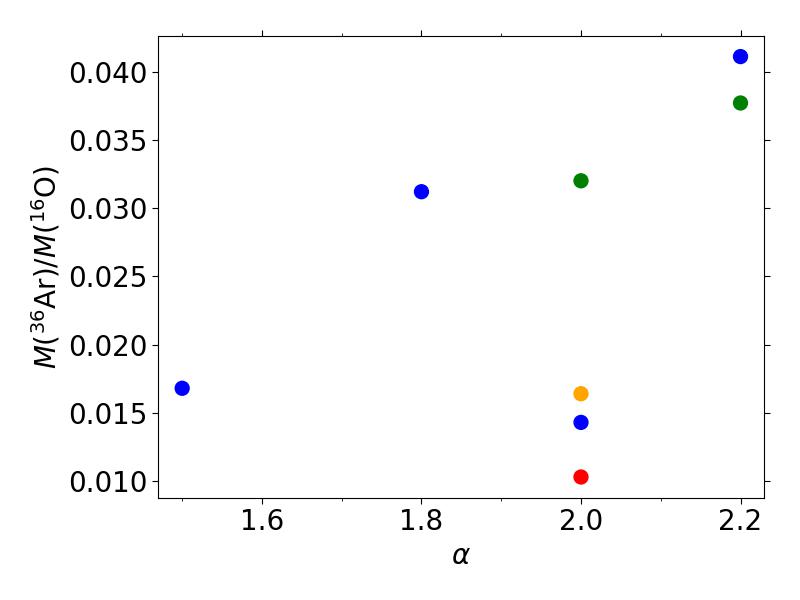}
    \includegraphics[width=8.5cm]{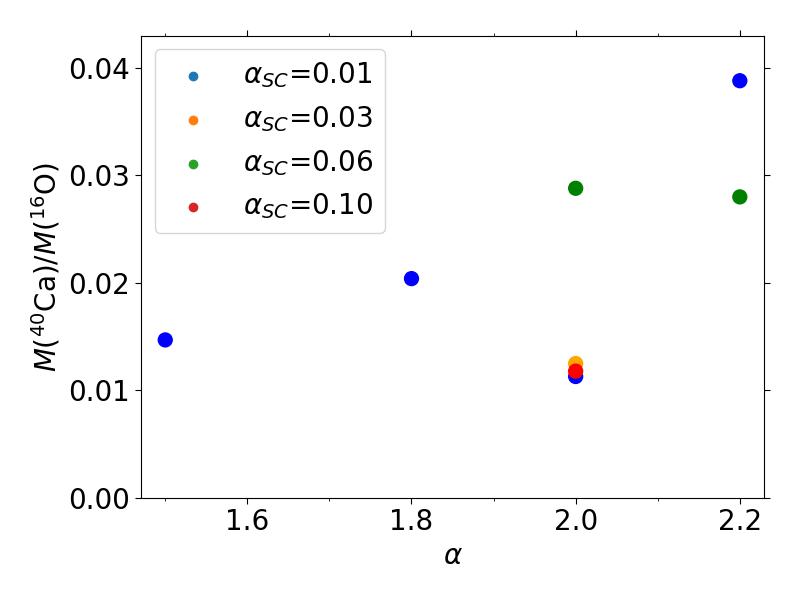}
    
    \caption{(top panel) The element mass ratio $M$(Si)/$M$(O) against $\alpha$ for all models for the 20 $M_{\odot}$ model. The colours (blue, green, orange, red) stand for $\alpha_{\rm SC} = 0.01, 0.03 0.06, 0.10$ respectively. (middle panel) Same as the top panel but for $M$(Ar)/$M$(O). (bottom panel) Same as the bottom panel but for $M$(Ca)/$M$(O).}
    \label{fig:xele_precollapse}
\end{figure}

The Fe-group elements are produced from the Si shell during the core-collapse supernova explosion. To further quantify the
production of these elements, we plot in Figure \ref{fig:xele_precollapse} the element mass ratios Si/O (top panel), $^{36}$Ar/$^{16}$O (middle panel) and Ca/O (bottom panel) for the models studied.  
A higher $\alpha$ model produces a higher production of Si/O mass ratio, while a higher $\alpha_{\rm SC}$ results in a lower Si/O ratio. From the Kippenhahn diagram and chemical profile, it confirms that $\alpha$ is primarily responsible for regulating the size of the Si-zone. A stronger convection allows more O to burn to form Si-group elements. For Ar/O and Ca/O, $\alpha$ and $\alpha_{\rm SC}$ have similar effects to their production in a similar mechanism. The rate of increase is different but the qualitative feature remains the same. One exception is M20A20S01 as noted above. Due to the interaction of convection zones in the Si and O layers, it results in a highly nonlinear behaviour compared to other models in the sequence. It has a drastically low Ar/O and Ca/O ratios. This agrees with the chemical profiles in Figure \ref{fig:precoll_profile_alpha}.

From the comparison with the Perseus Cluster, it is suggested that a high $\alpha$ is necessary to enhance the mass ratios of $^{36}$Ar/$^{16}$O and ${40}$Ca/${16}$O in the ejecta. At the same time, we expect that a high $\alpha_{\rm SC}$ is also necessary to suppress Si/O, so that it is not significantly overproduced compared to the Perseus Cluster. 

\section{Post-collapse Evolution}
\label{sec:post-collapse}

\subsection{Hydrodynamics}

When the massive star has begun its gravitational collapse, we transfer the stellar profile to the one-dimensional reactive hydrodynamics code. The core-collapse explosion is is initiated by setting the mass cut of the Fe core and the thermal-bomb type of energy deposition in the innermost $0.1~M_{\odot}$, which forms a shock wave propagating outward. Unless otherwise specified, we set the explosion energy, which is the final kinetic energy, as $10^{51}$ erg to match typical CCSN explosions. Also the time $t$ is measured from the onset of collapse. 

\begin{figure}
    \centering
    \includegraphics[width=8.5cm]{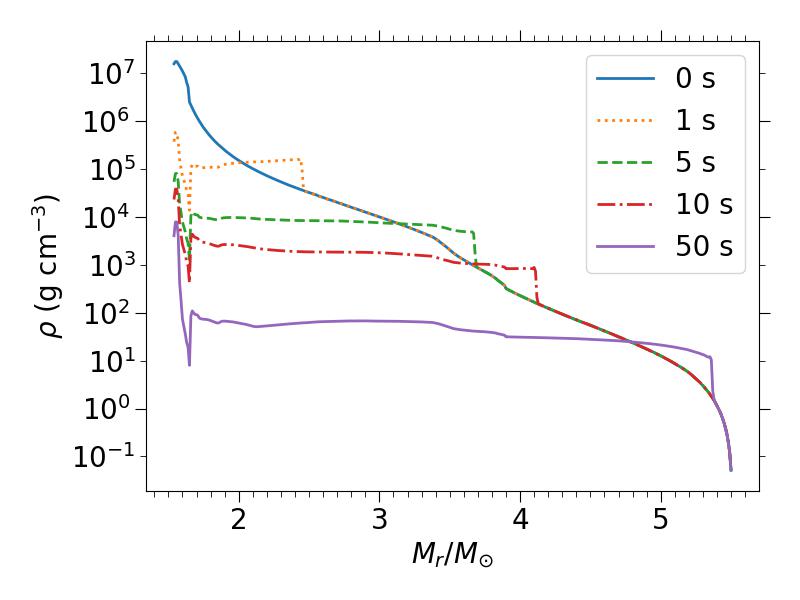}
    \caption{Density profile snapshots of M20A22S03 post-collapse explosion at $t = 0$ (blue solid line), 1 s (orange dotted line), 5 s (green dashed line), 10 s (red dot-dashed line), and 50 s (purple solid line). The time $t$ is measured from the onset of collapse.}
    \label{fig:hydro_rho}
\end{figure}

In Figure \ref{fig:hydro_rho}, we plot the density profile snapshot. We use M20A22S03 as a reference, since this model shows the progenitor chemical abundance patterns that are closest to the Perseus Cluster as described in previous sections. We choose to plot at $t = $ 0, 1, 5, 10 and 50 s respectively. The global profile shows a uniform expansion and the density of the plateau is monotonically decreasing. At the mean time, the density discontinuity gradually propagates outward. It starts from about $\sim 2~M_{\odot}$ and reaches the surface at about 50 s. 

\begin{figure}
    \centering
    \includegraphics[width=8.5cm]{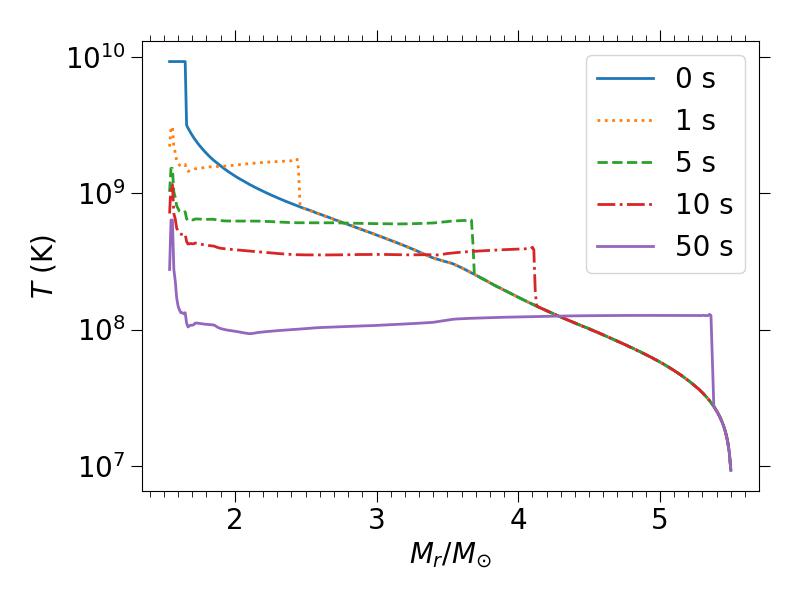}
    \caption{Same as Figure \ref{fig:hydro_rho} but for the temperature profile snapshots.}
    \label{fig:hydro_temp}
\end{figure}

Figure \ref{fig:hydro_temp} is analogous to Figure \ref{fig:hydro_rho}, showing instead the temperature profile snapshots of the same model. Due to radiation and thermal pressure, the temperature is more uniformly distributed at the plateau. At the end of the simulation, the temperature drops to about $10^8$ K. At such temperature, most exothermic reactions have ceased. 

\begin{figure}
    \centering
    \includegraphics[width=8.5cm]{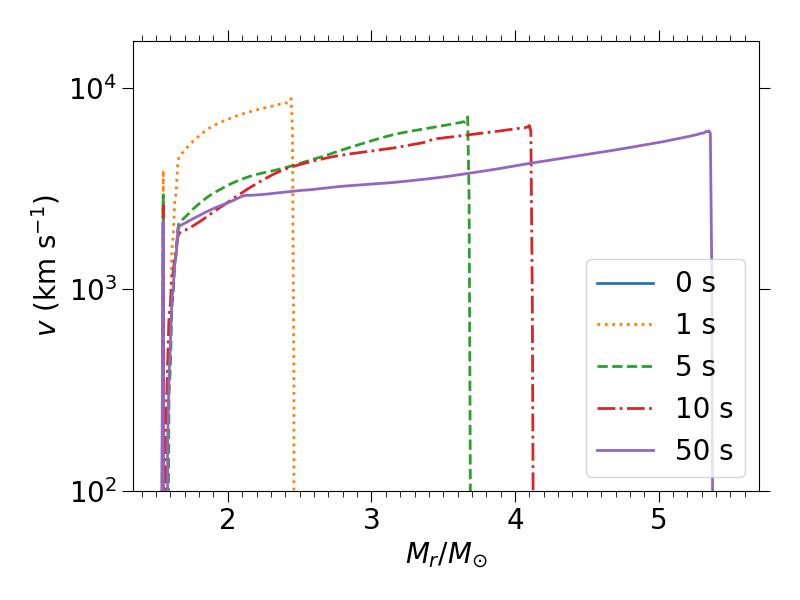}
    \caption{Same as Figure \ref{fig:hydro_rho} but for the velocity profile snapshots.}
    \label{fig:hydro_vel}
\end{figure}

We also plot the velocity profile in Figure \ref{fig:hydro_vel} to describe the kinematics of the ejecta. The shock propagation is more clearly captured through the velocity as we observe the ejecta is stationary exterior of the shock. Interior of the shock, the matter is expanding with a velocity $\sim 3-5 \times 10^3$ km s$^{-1}$. The expansion velocity mildly drops as the shock expands spherically outward. This agrees with the density profile where the negative gradient sustains the shock strength. 

\subsection{Effects on Thermodynamics Trajectory}

\begin{figure}
    \centering
    \includegraphics[width=8.5cm]{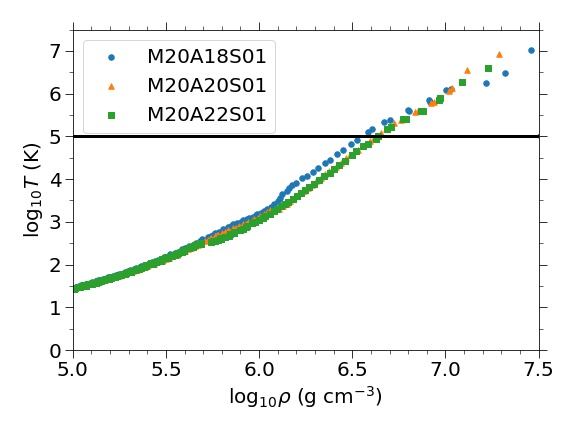}
    \includegraphics[width=8.5cm]{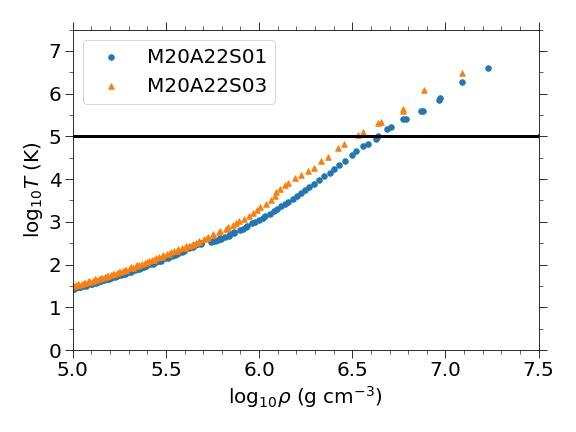} 
    \caption{(top panel) The corresponding density of the fluid elements at their maximum temperature taken from the explosive hydrodynamics simulations for M20A18S01 (blue circles), M20A20S01 (orange triangles) and M20A22S01 (green squares). The black solid line corresponds to the qualitative transition from incomplete burning (below the line) and $\alpha$-rich freezeout (above the line). (bottom panel) Same as the top panel but for M20A22S01 (blue circles) and M20A22S03 (orange triangles).}
    \label{fig:thermo_ref}
\end{figure}

To further calibrate the origin of the differences in our models, we plot in Figure \ref{fig:thermo_ref} the thermodynamics history of the fluid elements. We search for the density of each fluid element at their maximum temperature. For less massive stars, this density corresponds to the initial density of the fluid element. However, for more massive stars, the fluid elements have expanded for a period of time before the shock waves arrive and excite the fluid elements. In those cases, the instantaneous density at the shock excitation characterizes better its nucleosynthetic output. 

In the top panel, we plot the thermodynamics history for M20A18S01 (blue circles), M20A20S01 (orange triangles) and M20A22S01 (green squares). The region for incomplete burning\footnote{Incomplete burning stands for the nuclear reactions proceeding only up to Si-group elements, while complete burning stands for that proceeding up to $^{56}$Ni.}, and alpha-rich freezeout\footnote{$\alpha$-rich freezeout corresponds to the capture of the $\alpha$ particles by various nuclei during the ejecta expansion. These $\alpha$ particles come from photo-disintegration when the ejecta is hot $(> 5\times 10^9$ K). When the ejecta cools down, some $\alpha$ particles are captured by neighbouring nuclei, but the slow capture rate of $\alpha$ leaves behind an observable amount of $\alpha$ nuclei. The process is not in equilibrium and hence the composition contains less common isotopes including $^{44}$Ti and $^{56}$Co. See e.g., \cite{Jordan2003AlphaRich}.} are separated by the solid line.

When $\alpha$ is small, no observable change appears in the density and temperature 
upon shock arrival. 
Therefore, the change of the composition originates from the differences in the progenitor chemical composition. When $\alpha$ is increasing from 1.8 to 2.0, we observe a shift of the fluid element to a slightly higher density and temperature. This agrees with the upward trend of Ca/Fe and Ar/Fe production in previous plots. Further increasing $\alpha$ does not change the ignition density or temperature.

In the bottom panel of the same plot, we show the thermodynamics history for M20A22S01 (blue circles), M20A22S03 (orange triangles). Another shift is observed between the two models for the tracer with $\log_{10} \rho$ between 6.0 -- 6.5, stating that these tracers have experienced incomplete burning at a higher temperature but at a lower density. The two effects are balancing each other which results in a more uniform production of the elements. 

\subsection{Effects on Ejecta Profile}

\begin{figure}
    \centering
    \includegraphics[width=8.5cm]{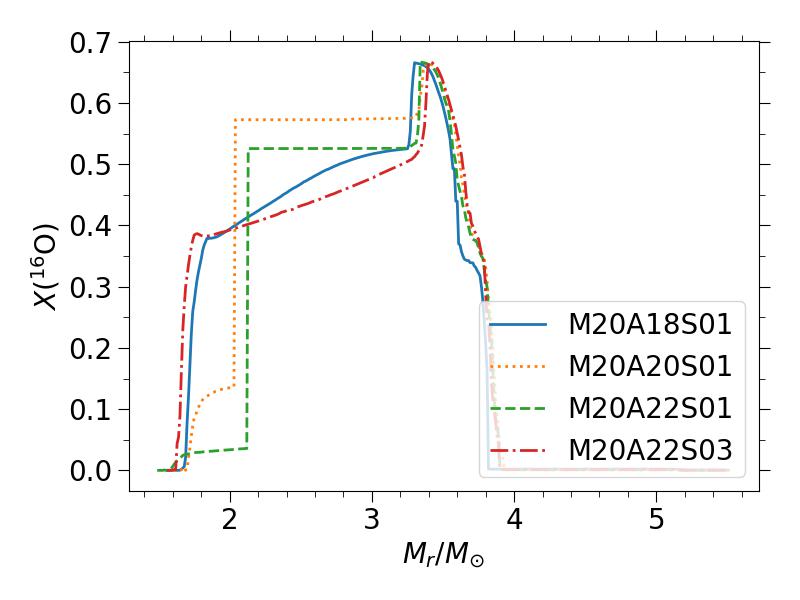}
    \includegraphics[width=8.5cm]{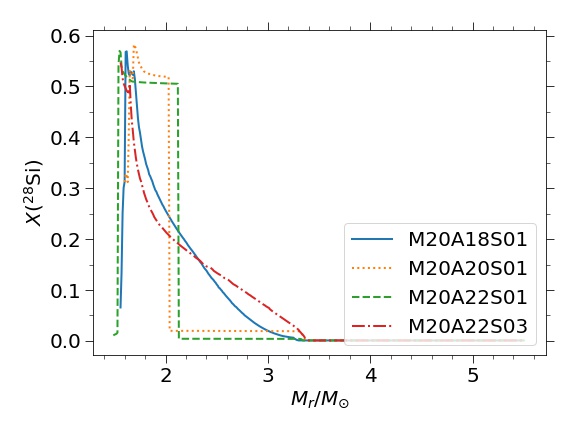}
    \includegraphics[width=8.5cm]{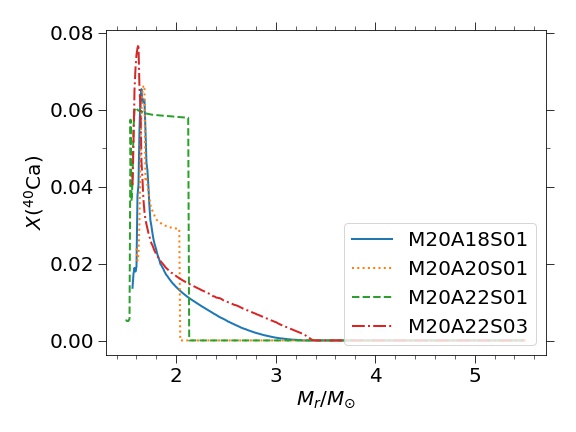}
        
    \caption{(top panel) The post-explosion $^{16}$O mass fraction in the ejecta for M20A18S01 (blue solid line), M20A20S01 (orange dotted line), M20A22S01 (green dashed line) and M20A22S03 (red dot-dashed line). (middle panel) Same as the top panel but for $^{28}$Si. (bottom panel) Same as the top panel but for $^{40}$Ca.}
    \label{fig:xiso_explo_ref}
\end{figure}

In Figure \ref{fig:xiso_explo_ref}, we plot the ejecta chemical profiles of selected elements for our model sequence. By comparing M20A18S01 (blue solid line), M20A20S01 (orange dotted line), and M20A22S01 (green dashed line), we observe that a higher $\alpha$ actually results in a more confined $^{16}$O distribution. On the other hand, the $^{28}$Si-rich zone is more clearly defined, which conforms with the higher $^{28}$Si production. The Ca-rich zone is also significantly extended when $\alpha$ becomes large. Therefore, comparing this and the previous figure, our results demonstrate that the increase in $\alpha$ enhances the Si-group element production 
because the progenitor contains a higher amount of these elements prior to the explosion.

By comparing M20A22S01 (green dashed line) and M20A22S03 (red dot-dashed line), the $\alpha_{\rm SC}$ shows an opposite effect on the ejecta distribution. The mass fraction profiles of $^{16}$O and $^{28}$Si are similar to M20A18S01. We emphasize that it does not mean the two parameters cancel each other. Globally, M20A22S03 has a larger amount of Si and Ca than M20A18S01. The two models have also very different pre-explosion profiles. 

The difference of the ejecta structure may be reflected in their velocity-space distributions. These models show that changing the convective mixing parameters can result in very different profiles of some elements showing long tails in the mass coordinates. When the photosphere recedes as the ejecta gradually becomes transparent, these elements will persist in the spectra, which may be compared with the stratified structure in the lower $\alpha$ or $\alpha_{\rm SC}$ models. As a result, future time series of the CCSN spectra can provide alternative constraints on these physical processes of mixing.

\section{Nucleosynthetic Yields}
\label{sec:catalogue}

\subsection{Effects on Isotopic Production}

With the explosive hydrodynamics simulation completed, we pass the thermodynamics history of the ejecta to compute the post-processing nucleosynthesis using the 495-isotope network \citep{Timmes1999Torch}. This allows us to capture the detailed chemical production. 

\begin{figure}
    \centering
    \includegraphics[width=8.5cm]{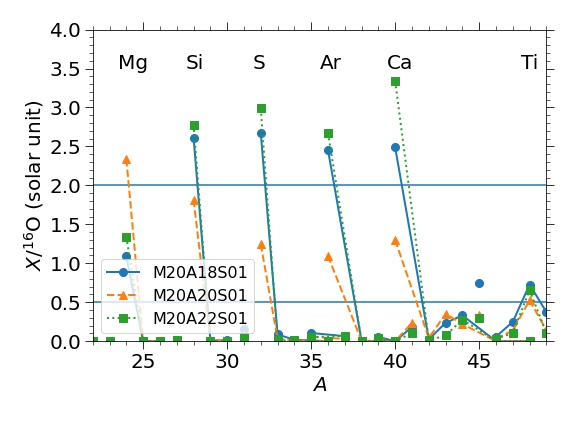}
    \includegraphics[width=8.5cm]{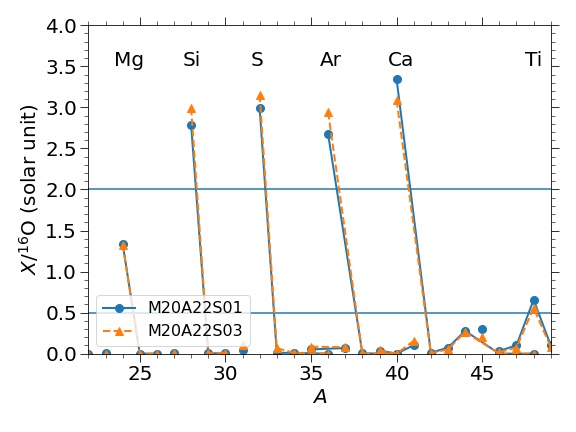} 
    \caption{(top panel) 
    The isotope ratio X/$^{16}$O for M20A18S01 (blue circles), M20A20S01 (orange triangles) and M20A22S01 (green triangles).
    (bottom panel) Same as the top panel but for M20A22S01 (blue circles) and M20A22S03 (orange triangles).}
    \label{fig:xiso_reference}
\end{figure}

In the top panel of Figure \ref{fig:xiso_reference} we plot the isotope ratio $X/^{16}$O in solar unit for M20A18S01, M20A20S01 and M20A22S01. We use the $^{16}$O as base such that we highlight how the yields of the new models within the massive star explosion family. Our models show that the Si-group elements are robustly produced. A higher $\alpha$ leads to higher production of these elements, including Ar and Ca. 
The production ratios from Ar and Ca in M20A22S01 is higher, especially Ca, which is important for easing the underproduction of this element.

In the bottom panel of the same figure, we compare the nucleosynthetic yields of M20A22S01 and M20A22S03. A higher $\alpha_{\rm SC}$ slightly enhances the production of the Si-group elements and reduces the difference among the four elements. 
The mass fractions of these elements are more uniform at a high $\alpha_{\rm SC}$. This is also important for boosting the production of Ar and Ca, without overproducing Si and S at the meantime. 

We remind that the typical SN Ia models \citep[see e.g., ][]{Leung2018Chand, Leung2020SubChand} have their Si-group elements slightly underproduced, which appears in different progenitors and explosion channels. Thus, for massive star models, slight overproduction could supplement the deficiencies of these elements in SNe Ia. 

\subsection{Effects on Elemental Production}

\begin{figure}
    \centering
    \includegraphics[width=8.5cm]{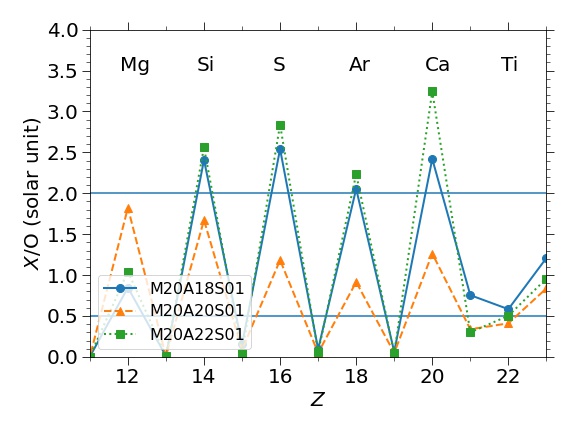}
    \includegraphics[width=8.5cm]{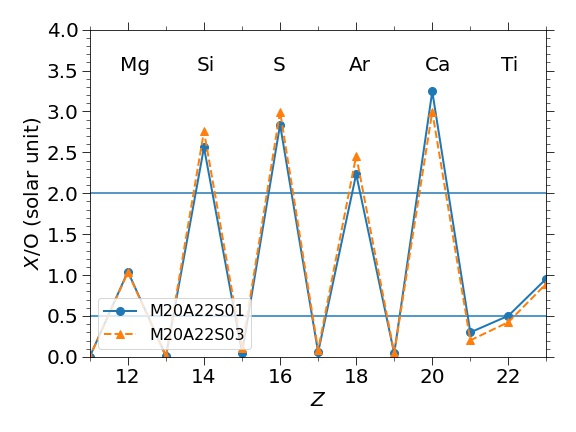} 
    \caption{(top panel) The element scaled mass fraction 
    X/O for M20A18S01 (blue circles) and M20A22S01 (orange triangles). The two horizontal lines correspond to 200 \% (top line) and 50 \% (bottom line) of the solar values.
    (bottom panel) Same as the top panel but for M20A22S01 (blue circles) and M20A22S03 (orange triangles).}
    \label{fig:xele_reference}
\end{figure}

In the top panel of Figure \ref{fig:xele_reference} we plot the elemental mass ratios Z/O
for M20A18S01, M20A20S01 and M20A22S01 to present the effects of $\alpha$ on elemental production. A higher $\alpha$ results in a higher production of the Si-group elements. Ca remains strongly enhanced. M20A18S01 and M20A22S02 show that these elements are above the 200\% line, suggesting that the choice of $\alpha$ does not change the role of CCSNe as the main producer of Si-group elements.

Similarly, in the bottom panel of Figure \ref{fig:xele_reference}, we plot the elemental mass ratios of M20A22S01 and M20A22S03 to contrast the role of $\alpha_{\rm SC}$. A higher $\alpha_{\rm SC}$ enhances production of the key elements, and the ratios among the four Si-group elements are closer to each other, which resembles with the patterns observed in the Perseus Cluster.

\subsection{Mass Dependence of Yields}

\begin{table*}[]
    \centering
    \caption{The massive star model sequence used in this work. $M_{\rm ZAMS}$, $M_{\rm ej}$ are the progenitor mass and ejected mass in units of $M_{\odot}$. $M(^{56}$Ni) is the mass of the ejected $^{56}$Ni in the explosion in units of $10^{-2}~M_{\odot}$. [Si/Fe], [S/Fe], [Ar/Fe], [Ca/Fe] are the element ratios of [Z/Fe] = $\log_{10}$ [(Z/Fe)/(Z/Fe)$_{\odot}$]. All models assumes the same explosion energy of $10^{51}$ erg.}
    \begin{tabular}{c c c c c c c c c c c c c}
        \hline
         Model & $M_{\rm ZAMS}$ & $\alpha$ & $\alpha_{\rm SC}$ & $M_{\rm He,C}$ & $M_{\rm C+O,C}$ & $M_{\rm Si,C}$ & $M_{\rm Fe,C}$ & $M(^{56}$Ni) & [Si/Fe] & [S/Fe] & [Ar/Fe] & [Ca/Fe] \\ \hline
         M15A22S03 & 15 & 2.2 & 0.03 & 3.68 & 2.17 & 1.68 & 1.36 & 9.86 & 0.013 & -0.17 & -0.24 & -0.040 \\
         M20A22S03 & 20 & 2.2 & 0.03 & 5.50 & 3.86 & 2.74 & 1.41 & 10.81 & 0.61 & 0.64 & 0.56 & 0.64 \\
         M25A22S03 & 25 & 2.2 & 0.03 & 7.34 & 7.22 & 1.69 & 1.40 & 10.25 & -0.085 & -0.22 & -0.28 & -0.096 \\
         M40A22S03 & 40 & 2.2 & 0.03 & 12.15 & 11.47 & 2.13 & 1.60 & 30.43 & -0.008 & -0.059 & -0.083 & 0.14 \\ \hline
    \end{tabular}
    
    \label{tab:models_mass}
\end{table*}

By examining how the explosive nucleosynthesis of the 20 $M_{\odot}$ progenitor depends on the convective parameter, it appears that the parameter set $\alpha = 2.2$ and $\alpha_{\rm SC} = 0.03$ has chemical abundance patterns which are the closest to the Perseus Cluster. In view of that, we extend our calculation using this set of parameters for various progenitor mass from $M_{\rm ZAMS} = 15 - 40~M_{\odot}$. In Table \ref{tab:models_mass}, we list the progenitor models and their explosive nucleosynthesis results. It is clear that a higher mass star has significantly larger C+O core mass. We also show the mass fraction ratios [Z/Fe] = $\log_{10}$ [(Z/Fe)/(Z/Fe)$_{\odot}$] for Si, S, Ar and Ca.

We note the following results:\\
\noindent (1) M15A22S03 and M25A22S03 both underproduce Si-group elements. \\
\noindent (2) M40A220S03 synthesizes the elements very close to the solar composition. \\
\noindent (3) M20A22S03 is the most prominent producer of Si group elements, with their abundance ratios a few times larger than the solar ratios. How similar the mass fraction ratios for Si, S, Ar, and Ca are also affects our interpretation of the observational data. \\
\noindent (4) M15A22S03 and M25A22S03 have lower production of S and Ar; while M20A22S03 has a uniform production over the four elements. \\
\noindent (5) In the Perseus cluster, the four elements are very close to solar values and their abundances are almost the same. 

\begin{figure}
    \centering
    \includegraphics[width=8.5cm]{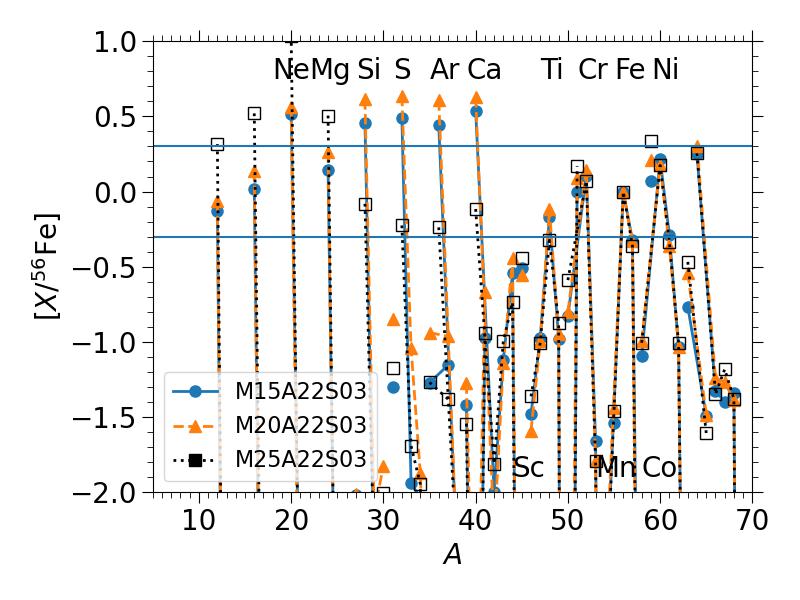}
    \includegraphics[width=8.5cm]{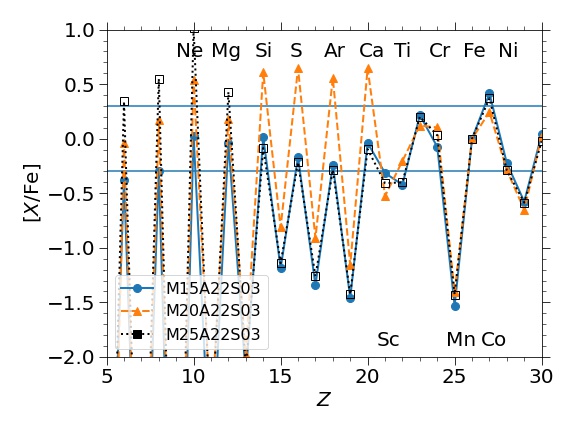} 
    \caption{(top panel) The isotopic abundance ratio for M15A22S03 (blue circles), M20A22S01 (orange triangles) and M25A22S03 (open black squares). The two horizontal lines correspond to 200\% (top line) and 50\%(bottom line) of the solar values.
    (bottom panel) Same as the top panel but for the elemental abundance ratio for the same sequences of models. Notice that the ratios of M25A22S03 are plotted in green squares.}
    \label{fig:xele_mass}
\end{figure}

We show in Figure \ref{fig:xele_mass} the chemical composition in the ejecta of $M_{\rm ZAMS} = 15, 20, 25~M_{\odot}$  in terms of isotopes (top panel) and elements (bottom panel). 
All three models show similar contributions to the Si-group elements. However, the more massive model (M25A22S03) has a higher contribution to the production of element Mg and odd-number elements such as P, Cl, K, and Sc. This is expected as these stars contain substantially larger amount of these elements in the progenitor envelopes. The middle-tier model (M20A22S03) has more influence on the Si-group elements, but it does not produce sufficient odd-number and neutron-rich isotopes (e.g., $^{29-30}$Si, $^{33-36}$S). Finally, M15A22S03 plays a secondary role in the light elements and similar role in Si-group elements compared to M20A22S03. 

In contrast, when we sum across the isotopes for the elemental production, it becomes clear that the 25 $M_{\odot}$ model primarily produces low-mass elements and 20 $M_{\odot}$ produces Si-group elements. And all three models show similar Fe-group element abundance patterns. This further conforms with our choice to take $20~M_{\odot}$ model as the reference model in the previous sections.

\section{Discussion}
\label{sec:discussion}

\subsection{Comparison with Literature}

As reported in \cite{Simionescu2019Perseus}, the canonical CCSN models do not perfectly fit the observational data of the Perseus Cluster, which comprise the most detailed constraints on the chemical composition of the ICM available to date. 
Here we compare our 20 $M_{\odot}$ CCSN model with those reported in the literature to contrast the differences in the chemical yield. Those models are also applied in that work to explain the Perseus Cluster and the 20 $M_{\odot}$ is the main contributor of the Si-group elements. 

\begin{figure}
    \centering
    \includegraphics[width=8.5cm]{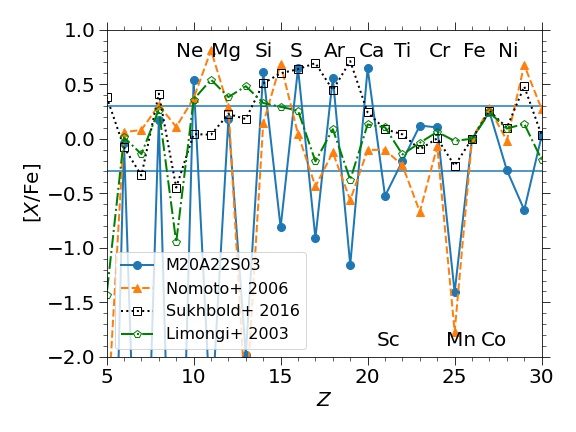}
    \caption{The 20 $M_{\odot}$ CCSN explosion models assuming solar metallcity taken from this work (M20A22S03, blue circles), N20 (orange triangles), S20 (open black squares), and L20 (green pentagons). }
    \label{fig:xele_literature}
\end{figure}

In Figure \ref{fig:xele_literature} we plot the chemical abundance pattern for M20A22S03 compared with two qualitatively different CCSN models, taken from \cite{Tominaga2007,  Sukhbold2016CCSN, Limongi2003CCSN} respectively to represent massive stars computed by different formalisms. The 20 $M_{\odot}$ model from \cite{Nomoto2006Yields} (denoted as N20) is built from the Henyey-type stellar evolution model reported in \cite{Nomoto1988StelEvol, Umeda2002PopIIICCSN, Ohkubo2009}. The explosion is triggered as a thermal bomb. A 240-isotope nuclear reaction network is used for the post-processing nucleosynthesis. The other 20 $M_{\odot}$ model from \cite{Sukhbold2016CCSN} (denoted as S20) uses the pre-supernova models calculated by the \texttt{Kepler} code \citep{Weaver1978Kepler, Woosley2015Kepler}. The explosion process is calculated by the Prometheus-Hot Bubble code \citep{Janka1996PTOBCode, Kifonidis2003Neutrino} and the code with a 2000-isotope network. At last, the 20 $M_{\odot}$ model \citep{Limongi2003CCSN}, denoted as L20, is computed by the \texttt{FRANEC} code with an adaptive network. A total of 44/149/267 isotopes are included for the H-/He-/advanced burning of the stars. The largest network involves elements from H to Mo. Ledoux criterion is used for convective mixing. Different from the previous two codes, the explosion can be modeled as a hydrodynamics effect, or in the radiation-dominated shock approximation.

The N20 model shows a similar even-odd elements parity to M20A22S03 where odd-number elements such as Al, P, Cl are remarkably lower than the neighbouring even-number elements such as Si, S, Ar. While the model shows a pronounced Na production, the Si-group elements are relatively close to solar value\footnote{We remind that SN Ia yields of Si-group elements are slightly sub-solar. Thus, in order to match the Perseus Cluster, super-solar yields of these elements in CCSNe are necessary.}. The model significantly underproduces Mn in comparison with the near-solar Ni yield. 

On the other hand, the S20 model shows a near-solar distribution for light elements (up to Al) and for iron-group elements from Ca onwards. Their high Si-group element production is similar to M20A22S03, while their model shows a smaller even-odd parity. This is likely a result of the large network where minor elements are tracked through the evolution. However, its Ca production is significantly lower than Si, S and Ar, which reduces its likeliness to directly match the Perseus cluster. S20 has also slightly sub-solar Mn and above-solar Ni yields. 

The L20 model has the overall the closest elemental distribution comparable to the solar composition. Some lower mass elements from Ne to S are overproduced. The Fe-group elements are effectively solar with the notable [Mn/Fe] being exactly solar, whereas all other models are showing sub-solar values. The large-network used in L20 also leads to a small difference between even-number and odd-number elements. On the other hand, the Si and S ratios relative to Fe are higher than Ar and Ca.

We remark that a direct comparison is difficult because the three models are obtained from calculations using different stellar evolution codes, input physics, and explosion mechanisms. As noted in this article, some processes such as convection can bring non-linear effects on the stellar evolution. Therefore, the differences could originate from both the differences in the stellar evolution and the explosive hydrodynamics. 
Further comparisons of other minor elements mentioned above, i.e., Sc, V and Zn, will provide additional insights on minor processes including the $\nu-p$ process \citep{Woosley1990NuP} and collapsar \citep{Tominaga2009Jet, Leung2023Jet1, Leung2024Jet2}. 

\subsection{Future Outlooks}

In this work, we have approximated the overall composition of supernova ejecta using only models with the solar progenitor metallcity. The Perseus Cluster, despite its gigantic structure and involvement of more than a thousand galaxies, shows a striking similar composition as the Sun. Thus, estimates using the solar metallicity may provide a first-hand estimation to the robustness of our new parameter choice. Meanwhile, if we need to consistently model the contribution of supernova ejecta to the hot gas in the Perseus Cluster over cosmic history, information about stars of various metallicity, and how the new models affects cumulatively generations of stars and their composition, will require elaborate galactic chemical evolution models \cite[see e.g.,][]{Mateteucci2001GCE, Kobayashi2020GCE}. Massive star models with a broader mass range and various progenitor metallicity will be an important extension of this work. 

Another shortcoming could be the network size used in this work. We follow previous approaches in constructing the massive star models and we use the thermal bomb to explode the star directly based on a small nuclear reaction network. This approach is accurate for predicting the even-number elements, including the elements studied in this work. However, reproducing odd-number elements, such as P, Al, K and Mn, will require consistent modeling of some minor elements. In order to consistently predict the production of these elements in massive stars, a larger network should be used to construct the detailed pre-explosion composition. Meanwhile, balance should be made such that each massive star model could be computed in reasonable time. How to extend the nuclear network will be an interesting follow-up project.

Third, to apply the new models in deriving the preferred supernova rates, the effect of Type Ia supernovae needs to be included consistently. While CCSNe produce a significant fraction of Si-group elements, a representative fraction comes from SNe Ia. Given the high precision measurement, combining the chemical yields of multiple astrophysical sources will yield a more reliable interpretation. 
Elements such as Mn are highly sensitive about the progenitor of Type Ia supernova \citep[e.g., ][]{Nomoto2017SNIa, Nomoto2018Review, Leung2021SN2014J}, to conclude if the supernovae are the explosion of the Chandrasekhar mass model \citep[e.g., ][]{Roepke2007DDT, Seitenzahl2013Chand, Leung2018Chand} and the sub-Chandrasekhar mass model \citep[e.g., ][]{Fink2010SubChand, Moll2013SubChand, Leung2020SubChand, Shen2021SubChand}.

Fourth, in this work we have removed the H-envelope in the advanced burning phase. It is possible that during the C- and O-burning, the high luminosity could continue to trigger mass loss, which directly remove the He-core, which could shift the evolutionary path of the star similar to those of lower mass by about $\sim5 \%$. This treatment could well approximate in binary systems where the H-envelope is removed by tidal interaction. However, for single stars, the He-core could be likely unchanged during the advanced phase for massive star $< 40~M_{\odot}$. The lowered He-core implies a more degenerate core at the point when the star collapses, which supports the production of Fe-group elements. It will be interesting to further contrast the differences in nucleosynthesis yield between unperturbed He-core (single star) and the He-core subject to mass loss (binary stars).

Lastly, in this work we have focused on the effects of the convective process to the production of Si-group elements. The convective process is one of the process which still face large uncertainties due to its multi-dimensional nature and the different preferred values in different stellar systems. It changes the yields by affecting the spread of elements in each layer, which changes the initial composition when the matter is burnt. Other processes, such as the $^{12}$C$(\alpha,\gamma)^{16}$O reaction rate, directly injects. However, such an approach could bring other impact. For example, the same rate is adjusted to explain the presence of black hole in the pair-instability mass gap \citep{Farmer2019Gap}. A systematic study is necessary to characterize the impact of this rate to various level of massive star evolution.

\subsection{Conclusions}

In this article we have studied the role of convection parameters, including the mixing length parameter $\alpha$ and the semi-convection parameter $\alpha_{\rm SC}$, in synthesizing Si-group elements (Si, S, Ar, and Ca). We provide in Appendix A the supernova yield table used in this work for matching the Perseus Cluster. We have used the 20 $M_{\odot}$ as a reference model to study their parameter dependence. Then we have studied their explosive nucleosynthesis to show how the differences in the progenitor chemical profile affects the final chemical yields of supernovae. We also have extended our calculations to various progenitor mass of $M_{\rm ZAMS} = 15 - 40~M_{\odot}$. We have applied our numerical results to match the observed chemical abundance of the Perseus Cluster. Below we list our discoveries reported in this article.

\begin{enumerate}
    \item The mixing length alpha $\alpha$ could lead to a more extended Si layer and mixing with the outer C+O layer. In some cases (such as M20A20S01) it could lead to a merger of convection zones in the Si and C+O-layers, which provides strong variations to the production of Ar and Ca. 

    \item The semi-convection parameter $\alpha_{\rm SC}$ also shifts the Si shell to the outer boundary, but has a small effect compared to $\alpha$.

    \item A higher $\alpha$ leads to a higher mass ratios for Si/O, S/O, Ar/O and Ca/O in the pre-collapse progenitor. Meanwhile, a higher $\alpha_{\rm SC}$ leads to lower even-odd mass ratios of these element pairs. The effects are the strongest for the Si-group elements. We conclude that $(\alpha, \alpha_{\rm SC}) = (2.2, 0.03)$ may provide the abundance pattern that is closest to the Perseus Cluster.

    \item We extend our models to $M_{\rm ZAMS} = $ 15 - 40 $M_{\odot}$ using the new parameter set result. The 20 $M_{\odot}$ progenitor is the main producer of Si-group elements compared to other stellar masses, while the 25 and 40 $M_{\odot}$ models strongly affect the overall production of light elements (C, O, Ne). All models show comparable contribution to the Fe-group elements. 

    
\end{enumerate}
 
\section*{Acknowledgment}

We thank Frank Timmes for the open-source subroutines of the Helmholtz equation of state and the torch nuclear reaction network. We also thank the MESA development team for making the MESA code open-source. 
This material is based upon work supported by the National Science Foundation under Grant AST-2316807.
K.N. acknowledges support by World Premier International Research Center Initiative (WPI), and JSPS KAKENHI Grant Numbers JP20K04024, JP21H04499, JP23K03452, and JP25K01046. 
A.S. acknowledges the Kavli IPMU for the continued hospitality. SRON Netherlands Institute for Space Research is supported financially by NWO.

\vspace{5mm}


\software{  Numpy \citep{Numpy},
            Matplotlib \citep{Matplotlib},
            Pandas \citep{Pandas}
          }

\appendix
\section{Supernova Yield Tables}

In the main text, we have presented the new massive star models for 15, 20, 25, and 40 $M_{\odot}$ stars. Here we provide the abundance yield table for the stable isotopes (Table \ref{table:yield_isotopes}) and long-lived unstable isotopes (Table \ref{table:unstable_isotopes}). We also compile the elemental yield in Table \ref{table:yield_elements} which lists the supernova yields of individual elements. 

For completeness, we also include the SN Ia Yields used in this work, including the Chandrasekhar mass model with DDT \citep{Leung2018Chand} and the sub-Chandrasekhar mass model with double detonation \citep{Leung2020SubChand}.

\begin{longtable}[c]{c c c c c c c c}
 \caption{The isotope yield table for supernova models. \label{table:yield_isotopes}} \\
 
 Isotope & M15A22S03 & M20A22S03 & M25A22S03 & M40A22S03 & Chand & sub-Chand (s) & sub-Chand (a) \\
 \hline
 \endfirsthead

 \multicolumn{8}{c}{\textit{Continuation of Table \ref{table:yield_isotopes}.}} \\
 Isotope & M15A22S03 & M20A22S03 & M25A22S03 & M40A22S03 & Chand & sub-Chand (s) & sub-Chand (a) \\
 \hline
 \endhead

 \hline
 \endfoot

 \hline
 \endlastfoot
 
$^{12}$C & $1.93 \times 10^{-1}$ & $3.81 \times 10^{-1}$ & $7.71 \times 10^{-1}
 $ & $15.82 \times 10^{-1}$ & $1.07 \times 10^{-3}$ & $1.20 \times 10^{-3}$ & $4
 .02 \times 10^{-3}$ \\
 $^{13}$C & $5.24 \times 10^{-10}$ & $3.87 \times 10^{-9}$ & $2.55 \times 10^{-9
 }$ & $4.77 \times 10^{-10}$ & $2.54 \times 10^{-12}$ & $3.18 \times 10^{-9}$ & 
 $8.97 \times 10^{-9}$ \\
 $^{14}$N & $1.44 \times 10^{-8}$ & $4.33 \times 10^{-8}$ & $4.88 \times 10^{-8}
 $ & $2.64 \times 10^{-7}$ & $1.40 \times 10^{-10}$ & $1.91 \times 10^{-8}$ & $1
 .17 \times 10^{-7}$ \\
 $^{15}$N & $1.70 \times 10^{-6}$ & $1.30 \times 10^{-6}$ & $1.21 \times 10^{-6}
 $ & $6.51 \times 10^{-6}$ & $9.40 \times 10^{-11}$ & $2.23 \times 10^{-10}$ & $
 1.38 \times 10^{-9}$ \\
 $^{16}$O & $5.07 \times 10^{-1}$ & $13.39 \times 10^{-1}$ & $27.29 \times 10^{-
 1}$ & $63.89 \times 10^{-1}$ & $5.69 \times 10^{-2}$ & $6.97 \times 10^{-2}$ & 
 $1.02 \times 10^{-1}$ \\
 $^{17}$O & $3.10 \times 10^{-11}$ & $3.18 \times 10^{-9}$ & $1.15 \times 10^{-8
 }$ & $1.20 \times 10^{-9}$ & $1.09 \times 10^{-11}$ & $1.13 \times 10^{-8}$ & $
 4.32 \times 10^{-8}$ \\
 $^{18}$O & $1.77 \times 10^{-11}$ & $9.96 \times 10^{-12}$ & $2.91 \times 10^{-
 11}$ & $1.05 \times 10^{-10}$ & $2.29 \times 10^{-13}$ & $9.91 \times 10^{-11}$
  & $5.10 \times 10^{-10}$ \\
 $^{19}$F & $1.08 \times 10^{-10}$ & $1.73 \times 10^{-10}$ & $1.27 \times 10^{-
 10}$ & $9.66 \times 10^{-10}$ & $1.38 \times 10^{-13}$ & $2.51 \times 10^{-11}$
  & $6.80 \times 10^{-11}$ \\
 $^{20}$Ne & $1.91 \times 10^{-1}$ & $5.61 \times 10^{-1}$ & $14.06 \times 10^{-
 1}$ & $23.30 \times 10^{-1}$ & $1.38 \times 10^{-4}$ & $1.21 \times 10^{-3}$ & 
 $4.72 \times 10^{-3}$ \\
 $^{21}$Ne & $1.44 \times 10^{-7}$ & $8.37 \times 10^{-8}$ & $1.24 \times 10^{-7
 }$ & $5.35 \times 10^{-7}$ & $3.06 \times 10^{-9}$ & $1.65 \times 10^{-7}$ & $4
 .33 \times 10^{-7}$ \\
 $^{22}$Ne & $2.20 \times 10^{-6}$ & $8.96 \times 10^{-7}$ & $7.42 \times 10^{-7
 }$ & $4.83 \times 10^{-6}$ & $4.28 \times 10^{-5}$ & $9.71 \times 10^{-9}$ & $9
 .46 \times 10^{-5}$ \\
 $^{23}$Na & $3.32 \times 10^{-6}$ & $1.04 \times 10^{-5}$ & $6.49 \times 10^{-6
 }$ & $3.32 \times 10^{-5}$ & $8.09 \times 10^{-7}$ & $8.81 \times 10^{-6}$ & $2
 .10 \times 10^{-5}$ \\
 $^{24}$Mg & $8.09 \times 10^{-2}$ & $1.20 \times 10^{-1}$ & $1.79 \times 10^{-1
 }$ & $3.44 \times 10^{-1}$ & $1.10 \times 10^{-3}$ & $1.34 \times 10^{-3}$ & $8
 .62 \times 10^{-3}$ \\
 $^{25}$Mg & $2.19 \times 10^{-7}$ & $3.59 \times 10^{-7}$ & $4.93 \times 10^{-7
 }$ & $1.20 \times 10^{-6}$ & $2.36 \times 10^{-6}$ & $1.52 \times 10^{-5}$ & $4
 .99 \times 10^{-5}$ \\
 $^{26}$Mg & $8.74 \times 10^{-6}$ & $3.34 \times 10^{-6}$ & $3.46 \times 10^{-6
 }$ & $1.54 \times 10^{-5}$ & $2.56 \times 10^{-6}$ & $3.03 \times 10^{-5}$ & $7
 .47 \times 10^{-5}$ \\
 $^{26}$Al & $9.56 \times 10^{-29}$ & $1.43 \times 10^{-28}$ & $2.23 \times 10^{
 -27}$ & $3.15 \times 10^{-28}$ & $3.56 \times 10^{-29}$ & $2.72 \times 10^{-29}
 $ & $2.86 \times 10^{-29}$ \\
 $^{27}$Al & $5.30 \times 10^{-5}$ & $7.38 \times 10^{-5}$ & $4.70 \times 10^{-5
 }$ & $1.77 \times 10^{-4}$ & $9.14 \times 10^{-5}$ & $1.20 \times 10^{-4}$ & $7
 .17 \times 10^{-4}$ \\
 $^{28}$Si & $9.93 \times 10^{-2}$ & $3.51 \times 10^{-1}$ & $5.99 \times 10^{-2
 }$ & $2.01 \times 10^{-1}$ & $2.35 \times 10^{-1}$ & $1.31 \times 10^{-1}$ & $1
 .10 \times 10^{-1}$ \\
 $^{29}$Si & $2.63 \times 10^{-5}$ & $3.30 \times 10^{-5}$ & $1.30 \times 10^{-5
 }$ & $2.40 \times 10^{-5}$ & $2.58 \times 10^{-4}$ & $2.80 \times 10^{-4}$ & $8
 .24 \times 10^{-4}$ \\
 $^{30}$Si & $3.86 \times 10^{-5}$ & $4.63 \times 10^{-5}$ & $2.56 \times 10^{-5
 }$ & $5.97 \times 10^{-5}$ & $3.51 \times 10^{-4}$ & $3.98 \times 10^{-4}$ & $1
 .58 \times 10^{-3}$ \\
 $^{31}$P & $4.48 \times 10^{-5}$ & $9.53 \times 10^{-5}$ & $3.78 \times 10^{-5}
 $ & $1.57 \times 10^{-4}$ & $1.92 \times 10^{-4}$ & $1.61 \times 10^{-4}$ & $3.
 54 \times 10^{-4}$ \\
 $^{32}$S & $3.72 \times 10^{-2}$ & $2.18 \times 10^{-1}$ & $2.53 \times 10^{-2}
 $ & $1.02 \times 10^{-1}$ & $1.23 \times 10^{-1}$ & $6.89 \times 10^{-2}$ & $4.
 90 \times 10^{-2}$ \\
 $^{33}$S & $7.76 \times 10^{-6}$ & $3.78 \times 10^{-5}$ & $7.16 \times 10^{-6}
 $ & $2.31 \times 10^{-5}$ & $2.85 \times 10^{-4}$ & $1.98 \times 10^{-4}$ & $3.
 08 \times 10^{-4}$ \\
 $^{34}$S & $4.19 \times 10^{-5}$ & $3.19 \times 10^{-5}$ & $2.29 \times 10^{-5}
 $ & $5.23 \times 10^{-5}$ & $2.09 \times 10^{-3}$ & $2.00 \times 10^{-3}$ & $1.
 97 \times 10^{-3}$ \\
 $^{36}$S & $3.07 \times 10^{-13}$ & $4.19 \times 10^{-11}$ & $3.70 \times 10^{-
 13}$ & $2.43 \times 10^{-12}$ & $3.35 \times 10^{-8}$ & $6.66 \times 10^{-8}$ &
  $1.82 \times 10^{-7}$ \\
 $^{35}$Cl & $2.43 \times 10^{-5}$ & $4.80 \times 10^{-5}$ & $1.89 \times 10^{-5
 }$ & $4.91 \times 10^{-5}$ & $1.53 \times 10^{-4}$ & $1.10 \times 10^{-4}$ & $1
 .30 \times 10^{-4}$ \\
 $^{37}$Cl & $1.78 \times 10^{-6}$ & $1.55 \times 10^{-5}$ & $5.02 \times 10^{-6
 }$ & $2.16 \times 10^{-6}$ & $5.07 \times 10^{-5}$ & $2.64 \times 10^{-5}$ & $2
 .37 \times 10^{-5}$ \\
 $^{36}$Ar & $7.61 \times 10^{-3}$ & $4.30 \times 10^{-2}$ & $5.20 \times 10^{-3
 }$ & $2.32 \times 10^{-2}$ & $2.22 \times 10^{-2}$ & $1.21 \times 10^{-2}$ & $7
 .52 \times 10^{-3}$ \\
 $^{38}$Ar & $1.05 \times 10^{-5}$ & $7.48 \times 10^{-6}$ & $6.30 \times 10^{-6
 }$ & $1.06 \times 10^{-5}$ & $1.82 \times 10^{-3}$ & $1.41 \times 10^{-3}$ & $1
 .04 \times 10^{-3}$ \\
 $^{40}$Ar & $1.95 \times 10^{-16}$ & $6.05 \times 10^{-14}$ & $3.07 \times 10^{
 -16}$ & $2.54 \times 10^{-15}$ & $8.26 \times 10^{-10}$ & $1.79 \times 10^{-9}$
  & $3.04 \times 10^{-9}$ \\
 $^{39}$K & $1.48 \times 10^{-5}$ & $2.23 \times 10^{-5}$ & $1.01 \times 10^{-5}
 $ & $2.21 \times 10^{-5}$ & $1.76 \times 10^{-4}$ & $9.71 \times 10^{-5}$ & $8.
 52 \times 10^{-5}$ \\
 $^{40}$K & $3.01 \times 10^{-13}$ & $9.92 \times 10^{-11}$ & $4.70 \times 10^{-
 13}$ & $3.72 \times 10^{-12}$ & $3.83 \times 10^{-8}$ & $4.03 \times 10^{-8}$ &
  $5.41 \times 10^{-8}$ \\
 $^{41}$K & $1.51 \times 10^{-6}$ & $7.14 \times 10^{-6}$ & $3.21 \times 10^{-6}
 $ & $2.00 \times 10^{-6}$ & $1.34 \times 10^{-5}$ & $5.92 \times 10^{-6}$ & $5.
 66 \times 10^{-6}$ \\
 $^{40}$Ca & $7.79 \times 10^{-3}$ & $3.40 \times 10^{-2}$ & $5.19 \times 10^{-3
 }$ & $2.50 \times 10^{-2}$ & $1.79 \times 10^{-2}$ & $1.09 \times 10^{-2}$ & $6
 .87 \times 10^{-3}$ \\
 $^{42}$Ca & $5.30 \times 10^{-7}$ & $2.97 \times 10^{-7}$ & $7.29 \times 10^{-7
 }$ & $1.77 \times 10^{-6}$ & $6.55 \times 10^{-5}$ & $4.45 \times 10^{-5}$ & $3
 .39 \times 10^{-5}$ \\
 $^{43}$Ca & $1.00 \times 10^{-6}$ & $9.43 \times 10^{-7}$ & $1.11 \times 10^{-6
 }$ & $3.05 \times 10^{-6}$ & $1.07 \times 10^{-6}$ & $1.45 \times 10^{-5}$ & $1
 .39 \times 10^{-5}$ \\
 $^{44}$Ca & $4.35 \times 10^{-5}$ & $6.77 \times 10^{-5}$ & $2.92 \times 10^{-5
 }$ & $8.86 \times 10^{-5}$ & $2.64 \times 10^{-5}$ & $2.76 \times 10^{-4}$ & $5
 .99 \times 10^{-4}$ \\
 $^{46}$Ca & $1.80 \times 10^{-23}$ & $9.66 \times 10^{-21}$ & $4.12 \times 10^{
 -23}$ & $1.46 \times 10^{-21}$ & $2.70 \times 10^{-11}$ & $5.14 \times 10^{-11}
 $ & $6.92 \times 10^{-11}$ \\
 $^{48}$Ca & $5.88 \times 10^{-25}$ & $5.28 \times 10^{-25}$ & $6.25 \times 10^{
 -25}$ & $1.16 \times 10^{-24}$ & $3.28 \times 10^{-14}$ & $9.74 \times 10^{-16}
 $ & $5.08 \times 10^{-16}$ \\
 $^{45}$Sc & $2.31 \times 10^{-6}$ & $1.28 \times 10^{-6}$ & $1.41 \times 10^{-6
 }$ & $2.42 \times 10^{-6}$ & $6.05 \times 10^{-7}$ & $3.96 \times 10^{-7}$ & $7
 .81 \times 10^{-7}$ \\
 $^{46}$Ti & $1.83 \times 10^{-6}$ & $7.39 \times 10^{-7}$ & $1.06 \times 10^{-6
 }$ & $1.57 \times 10^{-6}$ & $3.34 \times 10^{-5}$ & $1.94 \times 10^{-5}$ & $1
 .53 \times 10^{-5}$ \\
 $^{47}$Ti & $3.33 \times 10^{-6}$ & $2.76 \times 10^{-6}$ & $2.27 \times 10^{-6
 }$ & $5.92 \times 10^{-6}$ & $3.84 \times 10^{-6}$ & $2.39 \times 10^{-5}$ & $6
 .58 \times 10^{-5}$ \\
 $^{48}$Ti & $1.41 \times 10^{-4}$ & $2.15 \times 10^{-4}$ & $1.13 \times 10^{-4
 }$ & $4.71 \times 10^{-4}$ & $3.41 \times 10^{-4}$ & $7.44 \times 10^{-4}$ & $2
 .61 \times 10^{-3}$ \\
 $^{49}$Ti & $2.40 \times 10^{-6}$ & $2.40 \times 10^{-6}$ & $2.41 \times 10^{-6
 }$ & $4.00 \times 10^{-6}$ & $2.82 \times 10^{-5}$ & $1.54 \times 10^{-5}$ & $2
 .83 \times 10^{-5}$ \\
 $^{50}$Ti & $1.26 \times 10^{-23}$ & $4.59 \times 10^{-19}$ & $2.47 \times 10^{
 -23}$ & $6.81 \times 10^{-22}$ & $2.66 \times 10^{-6}$ & $6.38 \times 10^{-10}$
  & $7.12 \times 10^{-10}$ \\
 $^{50}$V & $2.69 \times 10^{-18}$ & $1.81 \times 10^{-14}$ & $2.69 \times 10^{-
 18}$ & $8.63 \times 10^{-17}$ & $1.71 \times 10^{-8}$ & $3.87 \times 10^{-9}$ &
  $3.86 \times 10^{-9}$ \\
 $^{51}$V & $7.28 \times 10^{-5}$ & $5.26 \times 10^{-5}$ & $5.34 \times 10^{-5}
 $ & $6.40 \times 10^{-5}$ & $9.50 \times 10^{-5}$ & $8.00 \times 10^{-5}$ & $2.
 81 \times 10^{-4}$ \\
 $^{50}$Cr & $3.00 \times 10^{-5}$ & $1.44 \times 10^{-5}$ & $1.97 \times 10^{-5
 }$ & $1.91 \times 10^{-5}$ & $4.96 \times 10^{-4}$ & $1.48 \times 10^{-4}$ & $1
 .09 \times 10^{-4}$ \\
 $^{52}$Cr & $1.86 \times 10^{-3}$ & $2.55 \times 10^{-3}$ & $1.81 \times 10^{-3
 }$ & $5.43 \times 10^{-3}$ & $8.04 \times 10^{-3}$ & $3.08 \times 10^{-3}$ & $2
 .80 \times 10^{-3}$ \\
 $^{53}$Cr & $1.75 \times 10^{-5}$ & $1.74 \times 10^{-5}$ & $1.44 \times 10^{-5
 }$ & $4.03 \times 10^{-5}$ & $1.00 \times 10^{-3}$ & $2.25 \times 10^{-4}$ & $1
 .56 \times 10^{-4}$ \\
 $^{54}$Cr & $8.71 \times 10^{-13}$ & $8.48 \times 10^{-13}$ & $2.54 \times 10^{
 -13}$ & $1.11 \times 10^{-12}$ & $7.34 \times 10^{-5}$ & $4.62 \times 10^{-8}$ 
 & $6.70 \times 10^{-8}$ \\
 $^{55}$Mn & $5.12 \times 10^{-5}$ & $6.14 \times 10^{-5}$ & $4.95 \times 10^{-5
 }$ & $1.52 \times 10^{-4}$ & $1.03 \times 10^{-2}$ & $1.38 \times 10^{-3}$ & $9
 .15 \times 10^{-4}$ \\
 $^{54}$Fe & $1.89 \times 10^{-5}$ & $1.91 \times 10^{-5}$ & $9.75 \times 10^{-6
 }$ & $3.06 \times 10^{-5}$ & $1.06 \times 10^{-1}$ & $9.12 \times 10^{-3}$ & $5
 .20 \times 10^{-3}$ \\
 $^{56}$Fe & $1.70 \times 10^{-1}$ & $1.53 \times 10^{-1}$ & $1.29 \times 10^{-1
 }$ & $3.62 \times 10^{-1}$ & $6.71 \times 10^{-1}$ & $6.30 \times 10^{-1}$ & $6
 .84 \times 10^{-1}$ \\
 $^{57}$Fe & $1.86 \times 10^{-3}$ & $1.74 \times 10^{-3}$ & $1.37 \times 10^{-3
 }$ & $4.73 \times 10^{-3}$ & $2.08 \times 10^{-2}$ & $1.68 \times 10^{-2}$ & $1
 .90 \times 10^{-2}$ \\
 $^{58}$Fe & $5.15 \times 10^{-11}$ & $5.91 \times 10^{-11}$ & $1.36 \times 10^{
 -10}$ & $2.86 \times 10^{-9}$ & $4.54 \times 10^{-4}$ & $1.43 \times 10^{-8}$ &
  $4.83 \times 10^{-8}$ \\
 $^{60}$Fe & $1.90 \times 10^{-25}$ & $1.96 \times 10^{-25}$ & $2.28 \times 10^{
 -24}$ & $1.26 \times 10^{-21}$ & $1.09 \times 10^{-10}$ & $3.20 \times 10^{-18}
 $ & $1.11 \times 10^{-18}$ \\
 $^{59}$Co & $1.15 \times 10^{-3}$ & $7.02 \times 10^{-4}$ & $7.90 \times 10^{-4
 }$ & $6.42 \times 10^{-4}$ & $8.86 \times 10^{-4}$ & $5.54 \times 10^{-4}$ & $6
 .17 \times 10^{-4}$ \\
 $^{58}$Ni & $7.73 \times 10^{-4}$ & $6.35 \times 10^{-4}$ & $5.20 \times 10^{-4
 }$ & $1.34 \times 10^{-3}$ & $6.35 \times 10^{-2}$ & $2.62 \times 10^{-2}$ & $3
 .06 \times 10^{-2}$ \\
 $^{60}$Ni & $4.84 \times 10^{-3}$ & $3.73 \times 10^{-3}$ & $3.13 \times 10^{-3
 }$ & $8.68 \times 10^{-3}$ & $1.12 \times 10^{-2}$ & $6.43 \times 10^{-3}$ & $6
 .41 \times 10^{-3}$ \\
 $^{61}$Ni & $6.46 \times 10^{-5}$ & $5.00 \times 10^{-5}$ & $4.49 \times 10^{-5
 }$ & $1.38 \times 10^{-4}$ & $2.66 \times 10^{-4}$ & $2.92 \times 10^{-4}$ & $2
 .48 \times 10^{-4}$ \\
 $^{62}$Ni & $4.38 \times 10^{-5}$ & $3.35 \times 10^{-5}$ & $2.99 \times 10^{-5
 }$ & $7.01 \times 10^{-5}$ & $1.88 \times 10^{-3}$ & $1.80 \times 10^{-3}$ & $1
 .83 \times 10^{-3}$ \\
 $^{64}$Ni & $1.04 \times 10^{-18}$ & $4.08 \times 10^{-18}$ & $1.71 \times 10^{
 -17}$ & $6.40 \times 10^{-16}$ & $1.21 \times 10^{-7}$ & $4.82 \times 10^{-14}$
  & $8.03 \times 10^{-13}$ \\
 $^{63}$Cu & $3.01 \times 10^{-5}$ & $2.23 \times 10^{-5}$ & $2.24 \times 10^{-5
 }$ & $1.37 \times 10^{-5}$ & $1.77 \times 10^{-6}$ & $2.63 \times 10^{-6}$ & $2
 .24 \times 10^{-6}$ \\
 $^{65}$Cu & $1.08 \times 10^{-6}$ & $1.18 \times 10^{-6}$ & $7.52 \times 10^{-7
 }$ & $2.60 \times 10^{-6}$ & $2.36 \times 10^{-6}$ & $7.99 \times 10^{-6}$ & $3
 .49 \times 10^{-6}$ \\
 $^{64}$Zn & $3.01 \times 10^{-4}$ & $2.61 \times 10^{-4}$ & $1.97 \times 10^{-4
 }$ & $4.38 \times 10^{-4}$ & $1.81 \times 10^{-5}$ & $2.84 \times 10^{-5}$ & $1
 .30 \times 10^{-5}$ \\
 $^{66}$Zn & $4.36 \times 10^{-6}$ & $4.39 \times 10^{-6}$ & $2.91 \times 10^{-6
 }$ & $7.13 \times 10^{-6}$ & $2.72 \times 10^{-5}$ & $2.53 \times 10^{-5}$ & $2
 .15 \times 10^{-5}$ \\
 $^{67}$Zn & $7.61 \times 10^{-7}$ & $6.14 \times 10^{-7}$ & $6.33 \times 10^{-7
 }$ & $1.77 \times 10^{-7}$ & $3.42 \times 10^{-8}$ & $3.58 \times 10^{-7}$ & $1
 .21 \times 10^{-6}$ \\
 $^{68}$Zn & $2.80 \times 10^{-6}$ & $2.16 \times 10^{-6}$ & $1.86 \times 10^{-6
 }$ & $4.49 \times 10^{-6}$ & $2.52 \times 10^{-8}$ & $3.89 \times 10^{-7}$ & $4
 .07 \times 10^{-7}$ \\
 $^{70}$Zn & $4.48 \times 10^{-27}$ & $5.90 \times 10^{-27}$ & $6.71 \times 10^{
 -27}$ & $2.53 \times 10^{-25}$ & $2.68 \times 10^{-15}$ & $2.07 \times 10^{-18}
 $ & $5.22 \times 10^{-17}$ \\

 \end{longtable}

\begin{longtable}[c]{c c c c c c c c}
 \caption{The unstable isotope yield table for supernova models. \label{table:unstable_isotopes}} \\
 
 Isotope & M15A22S03 & M20A22S03 & M25A22S03 & M40A22S03 & Chand & sub-Chand (s) & sub-Chand (a) \\
 \hline
 \endfirsthead

 \multicolumn{8}{c}{\textit{Continuation of Table \ref{table:unstable_isotopes}.}} \\
 Isotope & M15A22S03 & M20A22S03 & M25A22S03 & M40A22S03 & Chand & sub-Chand (s) & sub-Chand (a) \\
 \hline
 \endhead

 \hline
 \endfoot

 \hline
 \endlastfoot

   $^{22}$Na & $6.70 \times 10^{-8}$ & $5.32 \times 10^{-7}$ & $2.52 \times 10^{-7
 }$ & $6.75 \times 10^{-7}$ & $3.99 \times 10^{-10}$ & $3.92 \times 10^{-9}$ & $
 1.56 \times 10^{-8}$ \\
 $^{26}$Al & $1.04 \times 10^{-6}$ & $2.97 \times 10^{-6}$ & $3.12 \times 10^{-6
 }$ & $1.04 \times 10^{-5}$ & $3.45 \times 10^{-7}$ & $1.79 \times 10^{-6}$ & $8
 .12 \times 10^{-6}$ \\
 $^{39}$Ar & $4.92 \times 10^{-15}$ & $1.15 \times 10^{-12}$ & $6.03 \times 10^{
 -15}$ & $5.16 \times 10^{-14}$ & $5.00 \times 10^{-9}$ & $7.64 \times 10^{-9}$ 
 & $8.69 \times 10^{-9}$ \\
 $^{40}$K & $3.02 \times 10^{-13}$ & $9.97 \times 10^{-11}$ & $4.72 \times 10^{-
 13}$ & $3.74 \times 10^{-12}$ & $3.85 \times 10^{-8}$ & $4.06 \times 10^{-8}$ &
  $5.44 \times 10^{-8}$ \\
 $^{41}$Ca & $2.97 \times 10^{-7}$ & $7.18 \times 10^{-6}$ & $3.23 \times 10^{-6
 }$ & $5.10 \times 10^{-7}$ & $1.34 \times 10^{-5}$ & $5.94 \times 10^{-6}$ & $5
 .71 \times 10^{-6}$ \\
 $^{44}$Ti & $4.38 \times 10^{-5}$ & $6.88 \times 10^{-5}$ & $2.97 \times 10^{-5
 }$ & $8.98 \times 10^{-5}$ & $2.46 \times 10^{-5}$ & $2.78 \times 10^{-4}$ & $5
 .99 \times 10^{-4}$ \\
 $^{48}$V & $1.66 \times 10^{-9}$ & $8.51 \times 10^{-9}$ & $3.34 \times 10^{-9}
 $ & $2.13 \times 10^{-8}$ & $1.10 \times 10^{-7}$ & $2.06 \times 10^{-7}$ & $1.
 15 \times 10^{-6}$ \\
 $^{49}$V & $1.93 \times 10^{-10}$ & $7.07 \times 10^{-9}$ & $5.32 \times 10^{-9
 }$ & $9.19 \times 10^{-10}$ & $3.20 \times 10^{-7}$ & $1.62 \times 10^{-7}$ & $
 1.85 \times 10^{-7}$ \\
 $^{53}$Mn & $2.48 \times 10^{-8}$ & $3.22 \times 10^{-7}$ & $2.22 \times 10^{-7
 }$ & $3.06 \times 10^{-7}$ & $3.81 \times 10^{-4}$ & $9.50 \times 10^{-6}$ & $1
 .15 \times 10^{-5}$ \\
 $^{60}$Fe & $1.95 \times 10^{-24}$ & $2.01 \times 10^{-24}$ & $3.19 \times 10^{
 -23}$ & $1.83 \times 10^{-20}$ & $1.60 \times 10^{-9}$ & $4.82 \times 10^{-17}$
  & $1.62 \times 10^{-17}$ \\
 $^{56}$Co & $1.34 \times 10^{-6}$ & $3.20 \times 10^{-6}$ & $2.74 \times 10^{-6
 }$ & $1.90 \times 10^{-5}$ & $9.96 \times 10^{-5}$ & $9.38 \times 10^{-6}$ & $1
 .79 \times 10^{-5}$ \\
 $^{57}$Co & $1.95 \times 10^{-7}$ & $3.32 \times 10^{-7}$ & $3.65 \times 10^{-7
 }$ & $3.03 \times 10^{-6}$ & $1.19 \times 10^{-3}$ & $5.90 \times 10^{-6}$ & $9
 .87 \times 10^{-6}$ \\
 $^{60}$Co & $3.49 \times 10^{-17}$ & $1.73 \times 10^{-17}$ & $1.52 \times 10^{
 -16}$ & $4.97 \times 10^{-15}$ & $7.70 \times 10^{-8}$ & $1.23 \times 10^{-13}$
  & $6.59 \times 10^{-13}$ \\
 $^{56}$Ni & $1.70 \times 10^{-1}$ & $1.53 \times 10^{-1}$ & $1.29 \times 10^{-1
 }$ & $3.62 \times 10^{-1}$ & $6.27 \times 10^{-1}$ & $6.30 \times 10^{-1}$ & $6
 .84 \times 10^{-1}$ \\
 $^{57}$Ni & $1.85 \times 10^{-3}$ & $1.75 \times 10^{-3}$ & $1.37 \times 10^{-3
 }$ & $4.75 \times 10^{-3}$ & $1.95 \times 10^{-2}$ & $1.68 \times 10^{-2}$ & $1
 .90 \times 10^{-2}$ \\
 $^{59}$Ni & $8.54 \times 10^{-6}$ & $6.19 \times 10^{-5}$ & $5.86 \times 10^{-5
 }$ & $3.58 \times 10^{-5}$ & $4.05 \times 10^{-4}$ & $3.03 \times 10^{-6}$ & $4
 .77 \times 10^{-6}$ \\
 $^{63}$Ni & $2.18 \times 10^{-20}$ & $1.21 \times 10^{-19}$ & $3.22 \times 10^{
 -19}$ & $1.46 \times 10^{-17}$ & $5.61 \times 10^{-8}$ & $2.23 \times 10^{-15}$
  & $2.62 \times 10^{-14}$ \\

\end{longtable}

 \begin{longtable}[c]{c c c c c c c c}
 \caption{The elemental yield table for supernova models. \label{table:yield_elements}} \\
 
 Isotope & M15A22S03 & M20A22S03 & M25A22S03 & M40A22S03 & Chand & sub-Chand (s) & sub-Chand (a) \\
 \hline
 \endfirsthead

 \multicolumn{8}{c}{\textit{Continuation of Table \ref{table:yield_elements}.}} \\
 Isotope & M15A22S03 & M20A22S03 & M25A22S03 & M40A22S03 & Chand & sub-Chand (s) & sub-Chand (a) \\
 \hline
 \endhead

 \hline
 \endfoot

 \hline
 \endlastfoot

  C & $1.93 \times 10^{-1}$ & $3.81 \times 10^{-1}$ & $7.71 \times 10^{-1}$ & $15
 .82 \times 10^{-1}$ & $1.07 \times 10^{-3}$ & $1.20 \times 10^{-3}$ & $4.02 \times 10^{-3}$ \\
 N & $1.72 \times 10^{-6}$ & $1.34 \times 10^{-6}$ & $1.25 \times 10^{-6}$ & $6.
 77 \times 10^{-6}$ & $2.34 \times 10^{-10}$ & $1.93 \times 10^{-8}$ & $1.18 \times 10^{-7}$ \\
 O & $5.07 \times 10^{-1}$ & $13.39 \times 10^{-1}$ & $27.29 \times 10^{-1}$ & $
 63.89 \times 10^{-1}$ & $5.69 \times 10^{-2}$ & $6.97 \times 10^{-2}$ & $1.02 \times 10^{-1}$ \\
 F & $1.08 \times 10^{-10}$ & $1.73 \times 10^{-10}$ & $1.27 \times 10^{-10}$ & 
 $9.66 \times 10^{-10}$ & $1.38 \times 10^{-13}$ & $2.51 \times 10^{-11}$ & $6.8
 0 \times 10^{-11}$ \\
 Ne & $1.91 \times 10^{-1}$ & $5.61 \times 10^{-1}$ & $14.06 \times 10^{-1}$ & $
 23.30 \times 10^{-1}$ & $1.81 \times 10^{-4}$ & $1.21 \times 10^{-3}$ & $4.81 \times 10^{-3}$ \\
 Na & $3.32 \times 10^{-6}$ & $1.04 \times 10^{-5}$ & $6.49 \times 10^{-6}$ & $3
 .32 \times 10^{-5}$ & $8.09 \times 10^{-7}$ & $8.81 \times 10^{-6}$ & $2.10 \times 10^{-5}$ \\
 Mg & $8.09 \times 10^{-2}$ & $1.20 \times 10^{-1}$ & $1.79 \times 10^{-1}$ & $3
 .44 \times 10^{-1}$ & $1.11 \times 10^{-3}$ & $1.39 \times 10^{-3}$ & $8.75 \times 10^{-3}$ \\
 Al & $5.30 \times 10^{-5}$ & $7.38 \times 10^{-5}$ & $4.70 \times 10^{-5}$ & $1
 .77 \times 10^{-4}$ & $9.14 \times 10^{-5}$ & $1.20 \times 10^{-4}$ & $7.17 \times 10^{-4}$ \\
 Si & $9.94 \times 10^{-2}$ & $3.51 \times 10^{-1}$ & $5.99 \times 10^{-2}$ & $2
 .01 \times 10^{-1}$ & $2.35 \times 10^{-1}$ & $1.32 \times 10^{-1}$ & $1.13 \times 10^{-1}$ \\
 P & $4.48 \times 10^{-5}$ & $9.53 \times 10^{-5}$ & $3.78 \times 10^{-5}$ & $1.
 57 \times 10^{-4}$ & $1.92 \times 10^{-4}$ & $1.61 \times 10^{-4}$ & $3.54 \times 10^{-4}$ \\
 S & $3.72 \times 10^{-2}$ & $2.18 \times 10^{-1}$ & $2.53 \times 10^{-2}$ & $1.
 02 \times 10^{-1}$ & $1.25 \times 10^{-1}$ & $7.11 \times 10^{-2}$ & $5.13 \times 10^{-2}$ \\
 Cl & $2.61 \times 10^{-5}$ & $6.35 \times 10^{-5}$ & $2.39 \times 10^{-5}$ & $5
 .13 \times 10^{-5}$ & $2.04 \times 10^{-4}$ & $1.36 \times 10^{-4}$ & $1.54 \times 10^{-4}$ \\
 Ar & $7.63 \times 10^{-3}$ & $4.30 \times 10^{-2}$ & $5.21 \times 10^{-3}$ & $2
 .32 \times 10^{-2}$ & $2.41 \times 10^{-2}$ & $1.35 \times 10^{-2}$ & $8.57 \times 10^{-3}$ \\
 K & $1.63 \times 10^{-5}$ & $2.95 \times 10^{-5}$ & $1.34 \times 10^{-5}$ & $2.
 41 \times 10^{-5}$ & $1.89 \times 10^{-4}$ & $1.03 \times 10^{-4}$ & $9.09 \times 10^{-5}$ \\
 Ca & $7.83 \times 10^{-3}$ & $3.41 \times 10^{-2}$ & $5.22 \times 10^{-3}$ & $2
 .51 \times 10^{-2}$ & $1.80 \times 10^{-2}$ & $1.12 \times 10^{-2}$ & $7.52 \times 10^{-3}$ \\
 Sc & $2.31 \times 10^{-6}$ & $1.28 \times 10^{-6}$ & $1.41 \times 10^{-6}$ & $2
 .42 \times 10^{-6}$ & $6.05 \times 10^{-7}$ & $3.96 \times 10^{-7}$ & $7.81 \times 10^{-7}$ \\
 Ti & $1.49 \times 10^{-4}$ & $2.21 \times 10^{-4}$ & $1.19 \times 10^{-4}$ & $4
 .82 \times 10^{-4}$ & $4.10 \times 10^{-4}$ & $8.03 \times 10^{-4}$ & $2.72 \times 10^{-3}$ \\
 V & $7.28 \times 10^{-5}$ & $5.26 \times 10^{-5}$ & $5.34 \times 10^{-5}$ & $6.
 40 \times 10^{-5}$ & $9.50 \times 10^{-5}$ & $8.00 \times 10^{-5}$ & $2.81 \times 10^{-4}$ \\
 Cr & $1.91 \times 10^{-3}$ & $2.58 \times 10^{-3}$ & $1.85 \times 10^{-3}$ & $5
 .49 \times 10^{-3}$ & $9.62 \times 10^{-3}$ & $3.46 \times 10^{-3}$ & $3.07 \times 10^{-3}$ \\
 Mn & $5.12 \times 10^{-5}$ & $6.14 \times 10^{-5}$ & $4.95 \times 10^{-5}$ & $1
 .52 \times 10^{-4}$ & $1.03 \times 10^{-2}$ & $1.38 \times 10^{-3}$ & $9.15 \times 10^{-4}$ \\
 Fe & $1.72 \times 10^{-1}$ & $1.55 \times 10^{-1}$ & $1.30 \times 10^{-1}$ & $3
 .67 \times 10^{-1}$ & $7.99 \times 10^{-1}$ & $6.56 \times 10^{-1}$ & $7.08 \times 10^{-1}$ \\
 Co & $1.15 \times 10^{-3}$ & $7.02 \times 10^{-4}$ & $7.90 \times 10^{-4}$ & $6
 .42 \times 10^{-4}$ & $8.86 \times 10^{-4}$ & $5.54 \times 10^{-4}$ & $6.17 \times 10^{-4}$ \\
 Ni & $5.72 \times 10^{-3}$ & $4.45 \times 10^{-3}$ & $3.72 \times 10^{-3}$ & $1
 .02 \times 10^{-2}$ & $7.70 \times 10^{-2}$ & $3.47 \times 10^{-2}$ & $3.91 \times 10^{-2}$ \\
 Cu & $3.11 \times 10^{-5}$ & $2.35 \times 10^{-5}$ & $2.31 \times 10^{-5}$ & $1
 .63 \times 10^{-5}$ & $4.13 \times 10^{-6}$ & $1.06 \times 10^{-5}$ & $5.73 \times 10^{-6}$ \\
 Zn & $3.08 \times 10^{-4}$ & $2.68 \times 10^{-4}$ & $2.02 \times 10^{-4}$ & $4
 .50 \times 10^{-4}$ & $4.53 \times 10^{-5}$ & $5.45 \times 10^{-5}$ & $3.62 \times 10^{-5}$ \\

\end{longtable}
\section{Abundance Pattern with respect to Fe}

\begin{figure}
    \centering
    \includegraphics[width=8.5cm]{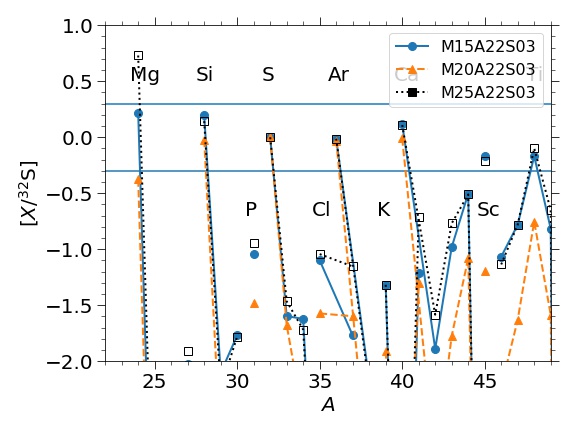}
    \includegraphics[width=8.5cm]{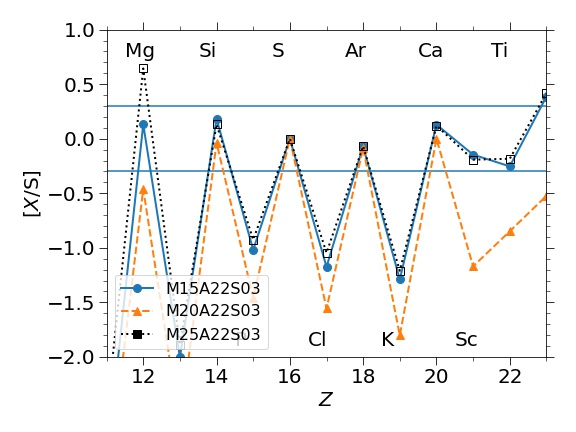} 
    \caption{(top panel) The isotopic abundance ratio with respect to $^{32}$S for M15A22S03 (blue circles), M20A22S01 (orange triangles) and M25A22S03 (open black squares). The two horizontal lines correspond to 200\% (top line) and 50\%(bottom line) of the solar values.
    (bottom panel) Same as the top panel but for the elemental abundance ratio to S for the same sequences of models. Notice that the ratios of M25A22S03 are plotted in green squares.}
    \label{fig:xele_mass2}
\end{figure}


In the main text we have presented results (Figure \ref{fig:xele_mass}) showing how our yield ratio X/Fe depends on different progenitor mass. 
Here we provide the additional plot of the abundance yield ratio with respect to S. We want to contrast the production mechanism in the Si-group elements, and also reduce the influence from SNe Ia.

In the upper panel of Figure \ref{fig:xele_mass2} we plot the isotopic abundance ratio [X/$^{32}$S] for our models with different progenitor masses. The ratio over $^{32}$S shows substantial difference to that over ${56}$Fe. Most even-number elements have similar ratios for different masses, which is in contrast with the mass ratios [X/Fe]. This is because ratio [X/S] directly compare the production of Si-group element in the same production site. Since the nuclear reaction channel does not change, locally the mass ratios remain the same. On the other hand, odd-number elements are more sensitive to the progenitor mass: A higher mass facilitates the production of these elements. For example, isotopes like $^{45}$Sc are closer to the solar value for the 15 and 25 $M_{\odot}$ models, compared to the 20 $M_{\odot}$ model.  

In the lower panel of the same figure we plot the elemental abundance ratio [X/S] for the same sequence of models. Our models show that all three models behave similarly. The even number elements (Si, S, Ar and Ca) are all very close to the solar values. The elemental behaviour for odd-number element in the 20 $M_{\odot}$ is also similar that it has a lower production in comparison with the other two.


\bibliographystyle{aasjournal}
\pagestyle{plain}
\bibliography{biblio}

\end{document}